\documentstyle[twoside]{article}

\catcode`\@=11
\long\def\@makefntext#1{
\protect\noindent \hbox to 3.2pt {\hskip-.9pt  
$^{{\eightrm\@thefnmark}}$\hfil}#1\hfill}               

\def\thefootnote{\fnsymbol{footnote}}
\def\@makefnmark{\hbox to 0pt{$^{\@thefnmark}$\hss}}    
        
\def\ps@myheadings{\let\@mkboth\@gobbletwo
\def\@oddhead{\hbox{}
\rightmark\hfil\eightrm\thepage}   
\def\@oddfoot{}\def\@evenhead{\eightrm\thepage\hfil
\leftmark\hbox{}}\def\@evenfoot{}
\def\sectionmark##1{}\def\subsectionmark##1{}}

\oddsidemargin=\evensidemargin
\renewcommand{\thefootnote}{\fnsymbol{footnote}}

\newcounter{sectionc}\newcounter{subsectionc}\newcounter{subsubsectionc}
\renewcommand{\section}[1] {\vspace{12pt}
\refstepcounter{sectionc}                
\setcounter{subsectionc}{0}\setcounter{subsubsectionc}{0}\noindent 
        {\tenbf\thesectionc. #1}\par\vspace{5pt}}
\renewcommand{\subsection}[1] {\vspace{12pt}
\refstepcounter{subsectionc}             
        \setcounter{subsubsectionc}{0}\noindent 
        {\bf\thesectionc.\thesubsectionc. {\kern1pt \bfit #1}}\par\vspace{5pt}}
\renewcommand{\subsubsection}[1] {\vspace{12pt}
\refstepcounter{subsubsectionc}          
        \noindent{\tenrm\thesectionc.\thesubsectionc.\thesubsubsectionc.
        {\kern1pt \tenit #1}}\par\vspace{5pt}}
\newcommand{\nonumsection}[1] {\vspace{12pt}\noindent{\tenbf #1}
        \par\vspace{5pt}}

\newcounter{appendixc}
\newcounter{subappendixc}[appendixc]
\newcounter{subsubappendixc}[subappendixc]
\renewcommand{\thesubappendixc}{\Alph{appendixc}.\arabic{subappendixc}}
\renewcommand{\thesubsubappendixc}
        {\Alph{appendixc}.\arabic{subappendixc}.\arabic{subsubappendixc}}

\renewcommand{\appendix}[1] {\vspace{12pt}
        \refstepcounter{appendixc}
        \setcounter{figure}{0}
        \setcounter{table}{0}
        \setcounter{lemma}{0}
        \setcounter{theorem}{0}
        \setcounter{corollary}{0}
        \setcounter{definition}{0}
        \setcounter{equation}{0}
        \renewcommand{\thefigure}{\Alph{appendixc}.\arabic{figure}}
        \renewcommand{\thetable}{\Alph{appendixc}.\arabic{table}}
        \renewcommand{\theappendixc}{\Alph{appendixc}}
        \renewcommand{\thelemma}{\Alph{appendixc}.\arabic{lemma}}
        \renewcommand{\thetheorem}{\Alph{appendixc}.\arabic{theorem}}
        \renewcommand{\thedefinition}{\Alph{appendixc}.\arabic{definition}}
        \renewcommand{\thecorollary}{\Alph{appendixc}.\arabic{corollary}}
        \renewcommand{\theequation}{\Alph{appendixc}.\arabic{equation}}
        \noindent{\tenbf Appendix \theappendixc #1}\par\vspace{5pt}}
\newcommand{\subappendix}[1] {\vspace{12pt}
        \refstepcounter{subappendixc}
        \noindent{\bf Appendix \thesubappendixc. {\kern1pt \bfit #1}}
        \par\vspace{5pt}}
\newcommand{\subsubappendix}[1] {\vspace{12pt}
        \refstepcounter{subsubappendixc}
        \noindent{\rm Appendix \thesubsubappendixc. {\kern1pt \tenit #1}}
        \par\vspace{5pt}}

\newcommand{\textlineskip}{\baselineskip=13pt}
\newcommand{\smalllineskip}{\baselineskip=10pt}

\def\eightcirc{
\begin{picture}(0,0)
\put(4.4,1.8){\circle{6.5}}
\end{picture}}
\def\eightcopyright{\eightcirc\kern2.7pt\hbox{\eightrm c}} 

\newcommand{\copyrightheading}[1]
        {\vspace*{-2.5cm}\smalllineskip{\flushleft
        {\footnotesize International Journal of Modern Physics B, #1}\\
        {\footnotesize $\eightcopyright$\, World Scientific Publishing
         Company}\\
         }}

\newcommand{\publisher}[2]{{\begin{center}\footnotesize\smalllineskip 
        Received #1\\
        Revised #2
        \end{center}
        }}

\def\abstracts#1#2#3{{
        \centering{\begin{minipage}{4.5in}\baselineskip=10pt\footnotesize
        \parindent=0pt #1\par 
        \parindent=15pt #2\par
        \parindent=15pt #3
        \end{minipage}}\par}} 
\def\abstracts#1#2#3#4#5#6#7#8#9{{
        \centering{\begin{minipage}{4.5in}\baselineskip=10pt\footnotesize
        \parindent=0pt #1\par 
        \parindent=15pt #2\par
        \parindent=15pt #3\par
        \parindent=15pt #4\par
        \parindent=15pt #5\par
        \parindent=15pt #6\par
        \parindent=15pt #7\par
        \parindent=15pt #8\par
        \parindent=15pt #9
        \end{minipage}}\par}} 

\renewenvironment{thebibliography}[1]                   
        {\frenchspacing
         \ninerm\baselineskip=11pt
         \begin{list}{\arabic{enumi}.}
        {\usecounter{enumi}\setlength{\parsep}{0pt}
         \setlength{\leftmargin 12.7pt}{\rightmargin 0pt} 
         \setlength{\itemsep}{0pt} \settowidth
        {\labelwidth}{#1.}\sloppy}}{\end{list}}

\newcounter{itemlistc}
\newcounter{romanlistc}
\newcounter{alphlistc}
\newcounter{arabiclistc}

\newcommand{\fcaption}[1]{
        \refstepcounter{figure}
        \setbox\@tempboxa = \hbox{\footnotesize Fig.~\thefigure. #1}
        \ifdim \wd\@tempboxa > 5in
           {\begin{center}
        \parbox{5in}{\footnotesize\smalllineskip Fig.~\thefigure. #1}
            \end{center}}
        \else
             {\begin{center}
             {\footnotesize Fig.~\thefigure. #1}
              \end{center}}
        \fi}

\newcommand{\tcaption}[1]{
        \refstepcounter{table}
        \setbox\@tempboxa = \hbox{\footnotesize Table~\thetable. #1}
        \ifdim \wd\@tempboxa > 5in
           {\begin{center}
        \parbox{5in}{\footnotesize\smalllineskip Table~\thetable. #1}
            \end{center}}
        \else
             {\begin{center}
             {\footnotesize Table~\thetable. #1}
              \end{center}}
        \fi}

\def\@citex[#1]#2{\if@filesw\immediate\write\@auxout
        {\string\citation{#2}}\fi
\def\@citea{}\@cite{\@for\@citeb:=#2\do
        {\@citea\def\@citea{,}\@ifundefined
        {b@\@citeb}{{\bf ?}\@warning
        {Citation `\@citeb' on page \thepage \space undefined}}
        {\csname b@\@citeb\endcsname}}}{#1}}

\newif\if@cghi
\def\cite{\@cghitrue\@ifnextchar [{\@tempswatrue
        \@citex}{\@tempswafalse\@citex[]}}
\def\citelow{\@cghifalse\@ifnextchar [{\@tempswatrue
        \@citex}{\@tempswafalse\@citex[]}}
\def\@cite#1#2{{$\null^{#1}$\if@tempswa\typeout
        {IJCGA warning: optional citation argument 
        ignored: `#2'} \fi}}

\def\@citexl[#1]#2{\if@filesw\immediate\write\@auxout
        {\string\citation{#2}}\fi
\def\@citeal{}\@citel{\@for\@citebl:=#2\do
        {\@citeal\def\@citeal{,}\@ifundefined
        {b@\@citebl}{{\bf ?}\@warning
        {Citation `\@citebl' on page \thepage \space undefined}}
        {\csname b@\@citebl\endcsname}}}{#1}}

\def\citelow{\@cghifalse\@ifnextchar [{\@tempswatrue
        \@citexl}{\@tempswafalse\@citexl[]}}
\def\@citel#1#2{{#1\if@tempswa\typeout
        {IJCGA warning: optional citation argument 
        ignored: `#2'} \fi}}
\def\pmb#1{\setbox0=\hbox{#1}
        \kern-.025em\copy0\kern-\wd0
        \kern.05em\copy0\kern-\wd0
        \kern-.025em\raise.0433em\box0}

\def\fnm#1{$^{\mbox{\scriptsize #1}}$}
\def\fnt#1#2{\footnotetext{\kern-.3em
        {$^{\mbox{\scriptsize #1}}$}{#2}}}

\def\fpage#1{\begingroup
\voffset=.3in
\thispagestyle{empty}\begin{table}[b]\centerline{\footnotesize #1}
        \end{table}\endgroup}

\def\runninghead#1#2{\pagestyle{myheadings}
\markboth{{\protect\footnotesize\it{\quad #1}}\hfill}
{\hfill{\protect\footnotesize\it{#2\quad}}}}
\font\tenrm=cmr10
\font\tenit=cmti10 
\font\tenbf=cmbx10
\font\bfit=cmbxti10 at 10pt
\font\ninerm=cmr9
\font\nineit=cmti9
\font\ninebf=cmbx9
\font\eightrm=cmr8

\def\qed{\hbox{${\vcenter{\vbox{                        
   \hrule height 0.4pt\hbox{\vrule width 0.4pt height 6pt
   \kern5pt\vrule width 0.4pt}\hrule height 0.4pt}}}$}}

\renewcommand{\thefootnote}{\fnsymbol{footnote}}        

\def\bsc{{\sc a\kern-6.4pt\sc a\kern-6.4pt\sc a}}       
\def\bflatex{\bf L\kern-.30em\raise.3ex\hbox{\bsc}\kern-.14em 
T\kern-.1667em\lower.7ex\hbox{E}\kern-.125em X} 

\newcommand{\beq}{\begin{equation}}
\newcommand{\eeq}{\end{equation}}
\newcommand{\barr}{\begin{eqnarray}}
\newcommand{\earr}{\end{eqnarray}}
\newcommand{\andy}[1]{ }

\def\h{\widehat}
\def\t{\widetilde}

\begin{document}

\runninghead{Temporal behavior of quantum mechanical systems}
 {Temporal behavior of quantum mechanical systems}

\normalsize\textlineskip
\thispagestyle{empty}
\setcounter{page}{1}

\copyrightheading{}                     

\vspace*{0.88truein}

\fpage{1}
\centerline{\bf TEMPORAL BEHAVIOR } 
\vspace*{0.035truein}
\centerline{\bf OF QUANTUM MECHANICAL SYSTEMS}
\vspace*{0.37truein}
\centerline{\footnotesize HIROMICHI NAKAZATO}
\vspace*{0.015truein}
\centerline{\footnotesize\it Department of Physics, Waseda University, 
3-4-1 Okubo, Shinjuku}
\baselineskip=10pt
\centerline{\footnotesize\it Tokyo 169, Japan}
\vspace*{10pt}
\centerline{\footnotesize MIKIO NAMIKI} 
\vspace*{0.015truein}
\centerline{\footnotesize\it Department of Physics, Waseda University, 
3-4-1 Okubo, Shinjuku}
\baselineskip=10pt
\centerline{\footnotesize\it Tokyo 169, Japan}
\vspace*{10pt}
\centerline{\normalsize and}
\vspace*{10pt}
\centerline{\footnotesize SAVERIO PASCAZIO}
\vspace*{0.015truein}
\centerline{\footnotesize\it Dipartimento di Fisica, Universit\`a di Bari, 
via Amendola 173}
\baselineskip=10pt
\centerline{\footnotesize\it and Istituto Nazionale di Fisica Nucleare, 
Sezione di Bari}
\baselineskip=10pt
\centerline{\footnotesize\it I-70126 Bari, Italy}
\vspace*{0.225truein}
\publisher{(received date)}{(revised date)}

\vspace*{0.21truein}
\abstracts{
The temporal behavior of quantum mechanical systems is reviewed.
We mainly focus our attention on the time development of the so-called 
``survival" probability of those systems that are 
initially prepared in eigenstates of the unperturbed Hamiltonian, 
by assuming that the latter has a continuous spectrum.}
{
The exponential decay of the survival probability, familiar, for
example, in radioactive decay phenomena, is representative of a purely
probabilistic character of the system under consideration and is naturally
expected to lead to a master equation.
This behavior, however, can be found only at intermediate times, for
deviations from it exist both at short and long times and can 
have significant consequences.}
{
After a short introduction to the long history of the research on
the temporal behavior
of such quantum mechanical systems, the short-time 
behavior and its controversial consequences when it is combined with 
von Neumann's projection postulate
in quantum measurement theory are critically
overviewed from a dynamical point of view. 
We also discuss the so-called quantum Zeno effect from this standpoint.}
{
The behavior of the survival amplitude is then scrutinized by 
investigating the analytic properties of its Fourier and Laplace transforms.
The analytic property that there is no singularity except a branch cut 
running along the real energy axis in the first Riemannian sheet is an 
important reflection of the time-reversal invariance of the dynamics 
governing the whole process.
It is shown that the exponential behavior is due to the presence of a simple 
pole in the second Riemannian sheet, while the contribution of the 
branch point yields a power behavior for the amplitude.
The exponential decay form is cancelled at short times and dominated at 
very long times by the branch-point contributions, which give a Gaussian 
behavior for the former and a power behavior for the latter.}
{
In order to realize the exponential law in quantum theory, it is essential 
to take into account a certain kind of macroscopic nature of the 
total system, since the exponential behavior
is regarded as a manifestation of a complete loss of coherence of the 
quantum subsystem under consideration.
In this respect, a few attempts at extracting the exponential decay form
on the basis of quantum theory, aiming at the master equation, are briefly 
reviewed, including van Hove's pioneering work and his
well-known ``$\lambda^2T$" limit.}
{
In the attempt to further clarify the mechanism of the appearance of a purely 
probabilistic behavior without resort to any approximation, a solvable 
dynamical model is presented and  extensively studied.
The model describes an ultrarelativistic
particle interacting with $N$ two-level systems (called ``spins") and
is shown to exhibit an exponential behavior at all times in the
weak-coupling, macroscopic limit.
Furthermore, it is shown that the model can even reproduce the short-time
Gaussian behavior followed by the exponential law when an
appropriate initial state is chosen.
The analysis is exact and no approximation is involved.
An interpretation for the change of the temporal behavior in quantum systems
is drawn from the results obtained.
Some implications for the quantum measurement problem are also discussed,
in particular in connection with dissipation.}{}{}{}

\textheight=7.8truein
\setcounter{footnote}{0}
\renewcommand{\thefootnote}{\alph{footnote}}
\noindent

\vspace*{1pt}\textlineskip      
\section{Introduction and summary}          
\vspace*{-0.5pt}
\label{sec-Intro}
\noindent
We know that the radioactive decay is well subject to the exponential
law, which can be easily understood on the basis of a purely 
probabilistic argument. We are also familiar with this kind of 
exponential behavior in a dissipative classical system, like for example
a one-particle motion characterized by {\em friction} 
in an atmosphere having a huge number of degrees of freedom. However,
an unstable nucleus in a radioactive material is usually in an excited 
quantum mechanical state, whose behavior is governed not by 
classical dynamics but by quantum mechanics. In this paper,
we review and examine what kind of temporal behavior should take 
place, and up to what extent we can observe the exponential decay 
in an unstable quantum mechanical system.

Let us first consider a classical system composed of many particles,
in which we are interested in a single-mode behavior 
(for example, a one-particle motion). On a fine time scale, 
whose unit is much smaller than the collision time,
we should observe a purely dynamical motion 
governed by the many-body Newton equations. On the other hand, if we observe 
the phenomenon on a macroscopic time scale, whose unit is 
much longer than the typical collision time, 
we can usually decompose the whole motion into two parts,
the first being a dissipative systematic motion, and the second
a fluctuating one, driven by a random force with a vanishing 
local time average. Here we are considering the local time average,
defined by an averaging procedure
over a local time region which is larger than the microscopic
relaxation time and smaller than the 
macroscopic time constant. 

In this view, we are dealing with 
the original many-body problem as a phenomenological 
stochastic process of a single-mode variable (for example, 
a one-particle motion), which is described
by the so-called {\em Langevin equation}. The systematic part, i.e.\
the local time average of the whole motion,
is usually subject to the exponential law, which is 
characterized by a {\em dissipation} constant (like the {\em friction}
constant in a one-particle motion). We can derive this description 
from the many-body Newton equations with the help of a 
coarse-graining procedure. When the number of dynamical degrees of 
freedom of the atmosphere is infinitely large, we can obtain the 
Gaussian distribution of the random force through the central
limit theorem, and then we can replace,
by means of a sort of {\em ergodic theorem},
the local time average with a statistical ensemble average
for the thermal equilibrium with a certain temperature $T$.  

On the other hand, when the microscopic
relaxation time is negligibly small, we can assume
that the random force is white and then the stochastic process 
becomes Markoffian. We usually call this kind of process a 
{\em Gauss-Markoffian} one. In this case, we can easily derive the
Fokker-Planck equation, which justifies the approach to thermal
equilibrium. Usually, the Langevin equation
is to be formulated for Gauss-Markoffian processes.
In this case, the ensemble average
of every physical quantity can be written as a sum of products 
of two-time correlation functions.
Note that we can also ascribe the process driven 
by a colored noise to a Gauss-Markoffian process, 
if the colored noise can be derived from a white noise 
through a simple mathematical transformation.

The situation with quantum systems is somewhat delicate for 
many reasons. In order to understand this situation more clearly, let us 
start our theoretical considerations from a general discussion 
on the temporal behavior of quantum-mechanical systems. 
First imagine, for example, that we have a particle system 
coupled with a field in free space. The case of a many-body
system with a finite (but very large) number of degrees of freedom 
will be discussed later.

Many years ago, Weisskopf and Wigner\cite{WW} first gave a 
simple formulation of the 
quantum theory of unstable systems, leading to the 
famous exponential law.  Breit and Wigner\cite{BW} introduced an 
expression for the quantum-mechanical wave functions or
$S$-matrix elements: Such an expression, which is analytic
with respect to the {\em energy} variable conjugated
to {\em time}, is equivalent to the exponential law.
Even though these formulas can work well
for phenomenological purposes, the theoretical 
procedure they gave was unfortunately not rigorous.
Another important analysis yielding the exponential law 
was given by Gamow in his seminal paper on the
quantum tunnelling problem.\cite{Gamow}

From the point of view of the relationship between the decay
phenomena and the energy-time uncertainty relation, Fock and 
Krylov\cite{FockKrylov} claimed that the exponential decay could not be 
theoretically accepted. Later, Khalfin\cite{Khalfin1957+58} 
confirmed this arguments on the basis of a mathematical theorem 
given by Paley and Wiener.\cite{PaleyWiener}

In 1953 Hellund\cite{Hell} strongly claimed that the long-time 
behavior is not subject to the exponential law, but rather to a power law.
By improving this conclusion, Namiki and Mugibayashi\cite{nm} showed 
that the quantum decay should be subject to the following 
three-step behavior: the Gaussian law at short times, 
the exponential law at intermediate times, and finally the power 
law at longer times. Later Araki, Munakata, Kawaguchi and Goto,\cite{ArakiMKG}
by making use of the Lee model,\cite{Lee}\
remarked that the exponential decay stems from simple poles located
on the second Riemannian sheet of the analytic expression of
the relevant $S$-matrix element. A 
similar but more thorough argument was also given by 
Schwinger.\cite{Schwinger}\ 
The quantum decay problem was also considered and reviewed 
by Fonda, Ghirardi and Rimini,\cite{FGR} and was recently discussed
by Cho, Kasari and Yamaguchi\cite{YY93b,YY94} by making 
use of a new solvable model of particle decay. Chiu, Sudarshan 
and Bhamathi\cite{Sudarshan2} and Horwitz\cite{Horwitz} presented other 
solvable models 
by extending the Lee model. One of the points of great interest in 
Refs.~\citelow{ArakiMKG} and \citelow{Sudarshan2} was the field-theoretical 
renormalization procedure. Throughout the present paper, however,
we shall not enter into the renormalization problem.

As for a general overview of the analytical
expressions, we can refer to early work on the damping theory
initiated by Heitler:\cite{Heitler}\ See, for example, the paper
by Arnous and Zienau.\cite{ArnousZienau}\ One should 
remember that the time-reversal invariance of the whole system
is reflected in the fact that the $S$-matrix element has no
singularity on the first Riemannian sheet, except the branch cut 
running along the real energy axis.

The above-mentioned temporal behavior of 
unstable quantum systems is closely related to the so-called
quantum Zeno paradox or quantum Zeno effect 
in the quantum measurement problem. The quantum Zeno paradox (QZP), 
named after the famous Greek philosopher Zeno, states that an 
unstable quantum system becomes stable (i.e.\ never decays) in the 
limit of infinitely frequent measurements. Of course, in practice, we cannot 
observe this very limit,\fnm{a}\fnt{a}{Even in principle, this limit cannot 
be observed, because of the uncertainty principle.
See Ref.~\citelow{qze2}.} but can only investigate the quantum Zeno 
effect (QZE), i.e.\ a milder version of the QZP, stating that the 
probability of finding the initial state is increased by a
(finite) number of repetitions of a measurement. We are now simply
considering the notion of quantum measurement in terms of the naive notion 
of wave function collapse (WFC), namely a simple projection onto the 
initial state.\cite{von}$^-$\cite{Bush}

The seminal idea of QZP or QZE was introduced under the assumption 
that the Gaussian short-time behavior can be observed and utilized in 
the relevant quantum decay and only the naive WFC takes place in 
quantum measurements.\cite{Beskow}$^-$\cite{Misra}\ In this 
context, one might think that the observation of this kind of 
phenomenon is a clear-cut experimental evidence in 
support of the naive WFC. 
Usually, the experimental observation of the Gaussian short-time behavior
in a decay process is so difficult 
to perform that Cook\cite{Cook} proposed 
to use atomic transitions of the oscillatory type, and inspired an  
important experiment\cite{Itano1} and then an interesting 
debate.\cite{Prigogine,qze1}\ 
As was discussed in the latter papers, we are led to the conclusion that the
experimental observation of the QZE does {\em not} necessarily support 
the naive WFC.\ \ In order to explain the situation in detail,
we have to examine one of the central questions in the quantum
measurement problem:
What is the wave-function collapse? In this context, the issue of 
the temporal development of quantum systems, by which we have obtained
the idea of QZP or QZE, is closely related to the quantum 
measurement problem. Two of the present authors have already
formulated a reasonable theory of quantum measurements without
resorting to the naive WFC.\cite{mn,np}\ \ In this paper, 
however, we shall not enter into this kind of problem.

Let us now consider and discuss a dynamical system composed of a finite number
of particles put in a finite box. The energy spectrum of this 
many-body system is discrete, and correspondingly, the 
elastic collision $S$-matrix element has a series of simple poles
running from the minimum energy (say, $E_{g}$) to infinity along 
the real axis on the complex $E$ plane. As the number of degrees of freedom 
and the box size become very (infinitely) large in an appropriate way,
the energy spectrum becomes asymptotically continuous and the series of simple
poles approaches a branch cut running from $E_{g}$ to infinity
along the real axis, just like in
the field-theoretical case mentioned above. This is a result of 
the asymptotic procedure, supplemented by an appropriate 
coarse-graining procedure, that endowes the off-diagonal matrix elements of 
the interaction Hamiltonian with random phases and then
provokes the appearance of the so-called 
{\em diagonal singularity}.\cite{vanHove}
A similar procedure was used in the theory of nuclear 
reactions.\cite{BlochNamiki}
As a consequence, the temporal behavior of the system is 
shown to be classified in 
three steps, the first being Gaussian, the second exponential, 
the third power-like.  Usually, the Gaussian period is very short and 
the power decay is very weak, so that the exponential decay seems to
dominate over the whole process. If we perform a sort of time-scale 
transformation in the weak-coupling case, following van Hove's 
procedure,\cite{vanHove} only the exponential decay part remains in 
the whole process. Being left with the exponential decay only, we can 
regard the whole process as a probabilistic one. Within the 
theoretical scheme given by van Hove, in fact, we can 
derive the master equation, which is a manifestation of irreversible 
processes, from the quantum-mechanical Schr\"{o}dinger equation.\cite{vanHove}

At the beginning of this introductory
section, we gave an outline of the derivation 
of the so-called Langevin equation from the many-body Newton equation 
in a classical system. Even in a quantum system, we can follow 
a similar kind of procedure by deriving a quantum Langevin equation
from the basic quantum-mechanical Schr\"{o}dinger or 
Heisenberg equations. Strictly speaking, however, we have to notice 
that all observables in quantum mechanics are represented by
operators, and their temporal evolution should obey an
operator-valued Heisenberg equation as a first principle. 
Therefore, we should first naturally derive an 
operator-valued Langevin equation. When and only when the operator
nature can be neglected, we can talk about a c-number Langevin 
equation as an approximation. Even in this case, the random force 
never rigorously satisfies the Gauss-Markoffian property, so that we cannot
calculate all the expectation values of the physical quantities by means 
of a sum of products of 
two-point correlation functions. When and only when 
the colored property of the quantum random force, and consequently all the 
important quantum properties, can be neglected, we can use a c-number 
quantum Langevin equation, which can be compared with the classical one.

If we want to formulate a quantum Langevin equation rigorously,
we also have to take into account the quantum properties of 
spin statistics, i.e.\ the well-known distinction between fermions 
and bosons. In order to do this, we should first formulate an 
operator-valued Langevin equation for quantum fields, as was already 
proposed by Mizutani, Muroya and Namiki.\cite{mmn}\ More details will be 
published in a forthcoming paper.

It is true that actual physical systems are so complicated that we cannot 
easily obtain exact solutions of the fundamental equation. In this context,
it is meaningful to examine the temporal evolution of 
quantum systems by means of exactly solvable models. 
In this paper, we shall describe in detail some 
attempts of this kind.\cite{NNP}$^-$\cite{NaPa3} This is one of the 
important purposes of the present paper.

We organize this paper in the following way. In Section 2, the short-time 
behavior of quantum mechanical systems is derived,
and its relations with the quantum Zeno paradox and 
quantum Zeno effect are critically discussed.
After a short introduction to the classical exponential law and 
the quantum mechanical deviations from it at short times, the 
difference between QZE and QZP is explained in Sec.~2.1.
The seminal formulation of the paradox is reviewed in 
Sec.~2.2, and an explicit example is considered in Sec.\ 2.3, where it
is also shown that both QZE and QZP can be given a purely dynamical
explanation. It is then shown in Sec.\ 2.4 that the 
QZP is only a mathematical idealization 
and is physically unattainable; however, its milder version, i.e.\ the 
quantum Zeno effect (QZE), can be observed in practice.
Section 3 deals with the long-time behavior of quantum systems. There,
the general expression for the survival probability amplitude is given 
for quantum systems with a huge (ideally infinite) 
number of degrees of freedom, where the
so-called random-phase approximation seems most plausible.
We see that the Fourier or Laplace transforms of the amplitude can clarify 
its analytical structure, which in turn determines its temporal behavior.
The following subsections aim at giving a deeper insight into the 
profound connection between the analytic structure and the time dependence 
of the amplitude. We also confirm the general statements on the basis of
a concrete, but still rather general Hamiltonian.
After a naive (textbook) derivation of the 
the short-time Gaussian behavior and the long-time exponential law,
a counter theorem claiming the unattainability of 
the exponential form at very long times is introduced in Sec.~3.1, 
and the analytical structure of the survival amplitude is
reinvestigated carefully and explicitly in Sec.~3.2.
It is shown that the exponential behavior is due to a simple pole located on
the second Riemannian sheet, and turns out to be dominated by a 
power decay at very long times. These exponential and power-like
 behaviors are a manifestation of the analytic 
property of the amplitude and the role played by the
branch cut in the complex 
energy plane is further analyzed in Sec.~3.3 on the basis of a 
concrete Hamiltonian. Section~4 is devoted to an explicit demonstration of the 
exponential decay form on the basis of a solvable dynamical model.
The model is solved exactly and a propagator, representative of the survival 
amplitude in this case, shows the exponential form at all times in the 
weak-coupling macroscopic limit.
The temporal behavior is shown to be sensitive to the choice of the initial 
state and the short-time Gassian as well as the 
following exponential behaviors 
are derived in the case of a wave-packet initial state.
A possible relation between the notions of dissipation and decoherence is 
mentioned, in relation to the quantum measurement problem.
The last section, Sec.\ 5, is devoted to conclusions and additional comments.


\section{Short time behavior and quantum Zeno effect}
\label{sec-stbeh}
\noindent
We shall start by giving a brief outline of the problem.
In classical physics, an expression for the decay 
probability of an unstable system is easily obtained by a heuristic 
approach (see, for example, Ref.\ \citelow{Gamow}):
One assumes that there is a {\em decay probability per unit time} $\Gamma$
that the system will decay according to a certain specific process.
Such a probability per unit time is constant, and does not depend, for 
instance, on the total number $N$ of unstable systems, on their past history, 
or on the environment surrounding them.
If the number of systems at time $t$ is $N(t)$ (a very large number),
the number of systems that will decay in the time interval $dt$ is
\andy{deccl}
\beq
dN = - N \Gamma dt \qquad \mbox{or} \qquad \frac{dN}{dt} = -\Gamma N ,
\label{eq:deccl}
\eeq
which yields 
\andy{decN}
\beq
N(t) = N_0 e^{-\Gamma t} ,
\label{eq:decN}
\eeq
where $N_0=N(0)$. One defines then the ``survival" or ``nondecay probability" 
\andy{decpri}
\beq
P(t) = \frac{N(t)}{N_0} = e^{-\Gamma t} .
\label{eq:decpri}
\eeq
The (positive) quantity $\Gamma$ is interpreted as the inverse of the 
``lifetime" $\tau$.
Notice the short-time expansion 
\andy{shti}
\beq
P(t) = 1 - \Gamma t + \cdots .
\label{eq:shti}
\eeq

Even though the above derivation can be found in elementary 
textbooks at undergraduate level, the assumptions underpinning it are delicate.
The reader will recognize the essential basic features of a
Markoffian process, in which memory effects are absent. More to this, one
is excluding {\em a priori} the possibility that cooperative effects take
place, making $\Gamma$ and $P$ environment-dependent. 
It is not surprising, then, that the resultant solution (\ref{eq:decpri}) 
exhibits all the basic ingredients of a dissipative behavior.

Let us now turn to a quantum mechanical description of the phenomenon.
Let $|a\rangle$ be the wave function of a quantum system $Q$ at time $t=0$.
The evolution of $Q$ is governed by the unitary operator
$U(t) = \exp (-iHt)$, where $H$ is the Hamiltonian.
We define the survival or nondecay probability at time $t$ as
the square modulus of the survival amplitude 
\andy{ndq}
\beq
P(t) = |\langle a|e^{-iHt}|a\rangle |^2.
\label{eq:ndq}
\eeq
A naive and elementary expansion at short times yields
\andy{naie}
\barr
P(t) & = & 1 - t^2 \left(\langle a|H^2|a\rangle - 
   ( \langle a|H|a\rangle )^2 \right) + \cdots  \nonumber \\
     & \equiv & 1 - t^2 \left( \triangle H \right)^2 + \cdots ,
\label{eq:naie}
\earr
which is quadratic in $t$ and therefore yields a vanishing decay rate for 
$t \rightarrow 0$. Observe that we are implicitly assuming that 
$|a\rangle$ be normalizable and 
$\left( \triangle H \right)^2$ nonvanishing, or, in other words, 
that $|a\rangle$ is not an eigenstate of $H$. 
(Note that one simply gets $P(t)=1$ if $|a\rangle$ is an eigenstate of $H$.) 
We also assumed that all moments of $H$ in the state $|a\rangle$
be finite.
Under these assumptions, we can easily infer that the survival probability
at short times is of the Gaussian type: \andy{Gauss}
\beq
P(t)\simeq \exp \left( -\frac{t^{2}}{\tau^{2}_{\rm G}} \right)\ 
:\quad \tau_{\rm G}^{-2}
\equiv (\Delta H)^{2}\ . \label{eq:Gauss}
\eeq
The temporal behavior of $P(t)$ is closely related to the famous issue of
the time-energy uncertainty relation, which has been a long-standing
argument of discussion.
In this context, we can refer to old papers, like for instance those by
Fock and Krylov\cite{FockKrylov} and Khalfin.\cite{Khalfin1968} 
The former claimed that the exponential decay was not  
theoretically acceptable,
and the latter confirmed this conclusion on the basis of a mathematical
theorem given by Paley and Wiener.\cite{PaleyWiener}

Needless to say, the above 
result is in manifest contradiction with the exponential law 
(\ref{eq:decpri}) that predicts a constant, nonvanishing decay rate
$\Gamma$. More to this,
there are some other important differences between the classical and the 
quantum mechanical results. First of all, observe that the derivation of 
(\ref{eq:ndq})-(\ref{eq:Gauss}) holds for individual systems, while
that of (\ref{eq:decpri}) makes use of an ensemble made up
of a huge number $N$ of identically prepared systems.
This is because probabilities are ontological in quantum mechanics, unlike 
in classical physics. 
Second, in the quantum case the initial state plays an important role 
and directly enters in the definition of $P$, raising subtle questions
about the problem of the initial state preparation, as is widely known:
See, for example, Refs.\ \citelow{vanHove} and \citelow{FGR}. 
By contrast, in the classical case, the initial state plays no crucial role,
and simply characterizes the constant $N_0$ in (\ref{eq:decN}).
This is obviously related to the underlying Markoffian assumptions.
Finally, notice that the quantum mechanical analysis holds true for any 
initial state $|a\rangle$, not necessarily unstable. 

The requirements that the initial state $|a \rangle$ be normalizable
and its energy finite are very important:
If these two conditions are not met, the behavior of 
the survival probability can be different from (\ref{eq:naie}) or 
(\ref{eq:Gauss}).
A typical example is the Breit-Wigner spectrum,
as was discussed in Refs.\ \citelow{Khalfin1957+58} and \citelow{lowbnd}.
The above two requirements will play an important role in 
the model we shall study in Sec.~\ref{sec-sdm}.

We shall postpone our considerations on the temporal behavior 
at longer times to
Sec.~\ref{sec-longt}. Now we analyze one of the most striking and 
famous consequences of the quadratic short-time behavior:
The quantum Zeno effect.

\subsection{Quantum Zeno paradox and quantum Zeno effect}
\label{sec-PE}
\noindent
Zeno was born in Elea, a southern Italian town not far from 
Naples, about 490 B.C. 
At those times, that region was part of the so-called {\em Magna Graecia}, 
and was culturally Greek.
Zeno was a disciple of Parmenides, the most prominent figure of 
the so-called Eleatic school of philosophers.
Parmenides firmly believed in a unique, entire, total {\em Truth} (``being"), 
and could not accept the idea that this Truth could change (``becoming")
or be composed of smaller entities (``multiplicity").
His dialogue with Socrates on these matters is one of the most famous
debates in the history of Greek philosophy.\cite{Russell}

Even though Zeno's contribution to the philosophy of ``being" were
not as important as Parmenides's, the former was an unequaled orator,
and is believed to have invented dialectic, that was subsequently 
so widely used by Socrates.
Two arguments given by Zeno in support of Parmenides's philosophical 
viewpoints are most famous. In the first one,
Achilles cannot reach a turtle because
when the former has arrived at the position previously occupied by 
the latter, the turtle has already moved away from it, and so on {\em ad
infinitum}.
In the second one, that concerns us more directly, a sped arrow
never reaches its target, because at every instant of time, by looking at the 
arrow, we clearly see that it occupies a definite position in space.
At every moment the arrow is therefore immobile, and 
by summing up so many ``immobilities" it is clearly impossible,
according to Zeno, to obtain motion.
If he lived nowadays, Zeno would probably consider the existence 
of photographs (in which all moving objects are still) as the best proof 
in support of his ideas.

In the light of his paradoxical arguments,
Zeno can be considered a forerunner of sophism. We know today that the two
above-mentioned paradoxes can be solved by infinitesimal
calculus. Nevertheless, one should not overlook the fact that Zeno 
basically aimed at giving very provocative arguments 
against the concept of ``becoming", in order to
ridicule the critics who tried to deride the philosophy of ``being".

It is somewhat astonishing that a conclusion similar to Zeno's
holds in quantum theory:
It is indeed possible to exploit the quantum mechanical vanishing 
decay rate at short times (\ref{eq:naie}) in a very interesting way,
by slowing down (and eventually halting) the decay process. 
The essential features of this effect were already known to 
von Neumann\cite{vNknew} and were investigated by several authors in the
past.\cite{Beskow}$^-$\cite{Misra}
Misra and Sudarshan first named this phenomenon after the Greek philosopher.

We shall first give an elementary derivation of this effect,
and present a more rigorous result in the following subsection.
Let us start from Eq.\ (\ref{eq:naie}) or (\ref{eq:Gauss}), 
that we rewrite as
\andy{sp1}
\beq
P(t) = 1 - \frac{t^2}{\tau^2_{\rm G}}+ \cdots 
\simeq e^{-t^2/\tau^2_{\rm G}},
\label{eq:sp1}
\eeq
where $\tau_{\rm G}$ is the characteristic time of the Gaussian evolution.
Suppose we perform $N$ measurements at equal time intervals $t$,
in order to ascertain whether the system is still in its initial state.
After each measurement, the system is ``projected" onto the quantum mechanical
state representing the result of the measurement, and the evolution starts 
anew. The total duration of the experiment is $T=Nt$.
The probability of observing the initial state at time $T$,
after having performed the $N$ above-mentioned measurements, reads 
\andy{spdy}
\beq
P^{(N)}(T) = [P(t)]^N =
  [P(T/N)]^N 
  \simeq \left( 1-\frac{1}{\tau^2_{\rm G}}
\left(\frac{T}{N} \right)^2 \right)^N 
\stackrel{N \; {\rm large}}{\sim}
e^{-T^2/\tau^2_{\rm G}N}.
\label{eq:spdy}
\eeq
Notice that both $T$ and $N$ are finite, in the above.
This is the {\em quantum Zeno effect}: Repeated observations
``slow down" the evolution and increase the probability that the system
is still in the initial state at time $T$.
In the limit of continuous observation ($N \rightarrow \infty$)
one obtains the {\em quantum Zeno paradox}
\andy{spdyN}
\beq
P^{(N)}(T) 
  \simeq \left( 1-\frac{1}{\tau^2_{\rm G}}
\left(\frac{T}{N} \right)^2 \right)^N 
\stackrel{N \rightarrow \infty}{\longrightarrow} 1.
\label{eq:spdyN}
\eeq
Infinitely frequent observations halt the evolution, and completely 
``freeze" the initial state of the quantum system.
Zeno seems to be right, after all: The quantum ``arrow", although sped
under the action of the Hamiltonian, 
does not move, if it is continuously observed.

What is the cause of this peculiar conclusion? Mathematically, the above 
result is ascribable to the general property
\andy{spdygen}
\beq
P^{(N)}(T) 
  \simeq \left( 1- O \left(\frac{1}{N^2} \right) \right)^N 
\stackrel{N \rightarrow \infty}{\longrightarrow} 1.
\label{eq:spdygen}
\eeq
However, the above limit is unphysical, for several reasons.
Indeed, notice that there is a profound difference between 
the $N$-finite and the $N$-infinite cases: 
In order to perform an experiment with $N$ finite one must 
only overcome practical problems, from the physical point of view.
(Of course, this can be a very difficult task, in practice.)
On the other hand, the $N \rightarrow \infty$ case 
is {\em physically unattainable}, and is rather to be regarded as a 
mathematical limit (although a very interesting one).
In this sense, we shall say that the quantum Zeno {\em effect},
with $N$ finite, becomes a quantum Zeno {\em paradox} when 
$N \rightarrow \infty$.\cite{qze1} 

It must also be emphasized that in the above analysis
the observations (measurements) are {\em instantaneous}, and this is
why it is possible to consider the $N \rightarrow \infty$
limit: No time is spent in measurements.\cite{qze1}
This is in line with the Copenhagen school of thought, and can be
regarded as a common and general feature of von Neumann-like 
descriptions of a measurement process.
Even though such a picture is often accepted among physicists,
it is very misleading, in our opinion, 
and hides some very important aspects of 
the problem: Indeed a measurement process, if analyzed (as it should)
as a concrete physical process,
takes place during a {\em very long time} on a microscopic scale, 
although we can regard it as if it happened {\em instantaneously}
on a macroscopic scale.

The physical unrealizability of the above-mentioned limit
will be thoroughly discussed in Sec.~\ref{sec-stbeh}.\ref{sec-Heiss},
where it will be shown that, as a consequence of the uncertainty relations, 
the $N \rightarrow \infty$ limit turns out to be impossible,
{\em even in principle}.\cite{qze2}

\subsection{Misra and Sudarshan's formulation of the quantum Zeno paradox}
\label{sec-MisSud}
\noindent
Let us now look at the seminal formulation of the quantum Zeno paradox
as given by Misra and Sudarshan.\cite{Misra} 
The derivation of the paradox is entirely based on von Neumann's projection
rule and therefore hinges upon the concept of quantum measurement process. The
reader should notice the fundamental role played by the projection operator 
${\cal O}$ in the following, and should remember that its effect  
on the state of the system is supposed to take place {\em instantaneously}.
We shall argue, in the next subsection, that the 
quantum Zeno effect and paradox can be derived on a purely 
dynamical basis, without making use of von Neumann's 
projections.\cite{qze1}

$Q$ is our unstable quantum system, whose states belong to the Hilbert
space ${\cal H}$ and whose evolution is described by the unitary 
operator $U(t) = \exp (-i H t)$, where $H$ is a semi-bounded Hamiltonian.
The initial density matrix of system $Q$ is assumed to be an undecayed state 
$\rho_0$, and let ${\cal O}$ be the projection operator over the subspace 
of the undecayed states. By definition,
\andy{inprep}
\beq
\rho_0 = {\cal O} \rho_0 {\cal O} , \qquad \mbox{Tr} [ \rho_0 {\cal O} ] = 1 .
\label{eq:inprep}
\eeq
Assume that we perform a measurement at time $t$, denoted by the projection 
operator ${\cal O}$, in order to check whether $Q$ is still undecayed. 
Accordingly,
the state of $Q$ changes into 
\andy{proie}
\beq
\rho_0 \rightarrow \rho(t) = {\cal O} U(t) \rho_0 U^\dagger(t) {\cal O} ,
\label{eq:proie}
\eeq
so that the probability of finding the system undecayed is given by
\andy{probini}
\beq
p(t) = \mbox{Tr} [ U(t) \rho_0 U^\dagger(t) {\cal O} ].
\label{eq:probini}
\eeq
The process (\ref{eq:proie}) will be referred to as ``naive wave function 
collapse". 
As already stressed, the observations (measurements) 
schematized via the operator ${\cal O}$ take place {\em instantaneously}.

The formulation of the QZP proceeds as
follows: We prepare $Q$ in the initial state $\rho_0$ at time 0 (this is
formally accomplished by performing an initial, ``preparatory" measurement of
${\cal O}$) and perform a series of observations at times $T/N, 2T/N, \ldots,
(N-1)T/N, T$. The state $\rho^{(N)}(T)$ of $Q$ after the preparation and the
above-mentioned $N$ measurements reads 
\andy{Nproie}
\beq
\rho^{(N)}(T) = V_N(T) \rho_0 V_N^\dagger(T) , \qquad  
    V_N(T) \equiv [ {\cal O} U(T/N) {\cal O} ]^N 
\label{eq:Nproie}
\eeq
and the probability of finding the system undecayed is given by
\andy{probNob}
\beq
P^{(N)}(t) = \mbox{Tr} \left[ V_N(T) \rho_0 V_N^\dagger(T) \right].
\label{eq:probNob}
\eeq
So far, the derivation is straightforward\fnm{b}\fnt{b}{Except, of course,
that one is led to wonder about the physical meaning of the operator 
${\cal O}$, and about the reason why a ``projection", unlike all other 
physical phenomena, should occur instantaneously. These questions are left 
to the reader. The von Neumann formalism was critically analyzed in
Refs.\ \citelow{mn} and \citelow{np}.} and does not involve, in our 
opinion, any new concept. By contrast, much more delicate is the set of 
assumptions leading to the ``paradox":
One considers the so-called limit of continuous observation, in which 
$N \rightarrow \infty$, and defines 
\andy{slim}
\beq
{\cal V} (T) \equiv \lim_{N \rightarrow \infty} V_N(T) ,
  \label{eq:slim}
\eeq
provided the above limit exists in the strong sense. 
The final state and the probability of observing the undecayed system 
read then
\andy{infproie,probinfob}
\barr
\t{\rho} (T) & = & {\cal V}(T) \rho_0 {\cal V}^\dagger (T),
  \label{eq:infproie} \\
{\cal P} (T) & \equiv & \lim_{N \rightarrow \infty} P^{(N)}(T) 
   = \mbox{Tr} \left[ {\cal V}(T) \rho_0 {\cal V}^\dagger(T) \right].
\label{eq:probinfob}
\earr
Moreover one assumes, on physical grounds, the 
strong continuity of ${\cal V}(t)$ in $t=0$,
\andy{phgr}
\beq
\lim_{t \rightarrow 0^+} {\cal V}(t) = {\cal O} .
\label{eq:phgr}
\eeq
Misra and Sudarshan then proved that under general 
conditions\fnm{c}\fnt{c}{Some technical requirements are needed for the proof.}
the operators ${\cal V}(T)$ (exist for all real $T$ and) form a semigroup
labeled by the time parameter $T$. Moreover,
${\cal V}^\dagger (T) = {\cal V}(-T)$, so that 
${\cal V}^\dagger (T) {\cal V}(T) ={\cal O}$.
This implies, by virtue of Eq.\ (\ref{eq:inprep}), that
\andy{probinfu}
\beq
{\cal P}(T) = \lim_{N \rightarrow \infty} P^{(N)}(T) = 
\mbox{Tr} \left[ \rho_0 {\cal O} \right] = 1 .
\label{eq:probinfu}
\eeq
If the particle is continuously observed (to check whether it decays or not),
it is ``frozen" in its initial state and will never be found to decay! 
This is the essence of the ``quantum Zeno paradox".

Two important problems are still open, at this point.
First: Is the present technique, that strongly hinges upon the action of 
projection operators {\em \`a la} von Neumann, really {\em necessary} 
in order to derive the QZP or the QZE?
Second: Is the $N \rightarrow \infty$ limit physically sensible?

We shall tackle the first question in Sec.~\ref{sec-stbeh}.\ref{sec-neutex},
where a purely dynamical derivation of the QZP and QZE is presented, that 
makes use of unitary evolutions and does not involve any projection 
(provided a final measurement is carried out).
We shall come back to the second question in 
Sec.~\ref{sec-stbeh}.\ref{sec-Heiss}, where the $N \rightarrow \infty$
limit will be shown to be in contradiction with the Heisenberg 
uncertainty principle.

\subsection{Dynamical quantum Zeno effect: A simple model}
\label{sec-neutex}
\noindent
Let us show that it is possible to otain both the QZE  
(\ref{eq:spdy}) or (\ref{eq:probNob}) and the QZP
(\ref{eq:spdyN}) or (\ref{eq:probinfu})
by making use of a purely dynamical process.
The general case is treated in Ref.\ \citelow{qze1}. Here, we
shall not give a general proof of the above statement, 
because the best proof is by inspection of 
an explicit example: Consider the experimental setup sketched in 
Figure~1.
\begin{figure}[htbp]
\vspace*{13pt}
\centerline{\vbox{\hrule width 5cm height0.001pt}}
\vspace*{2.8truein}             
\centerline{\vbox{\hrule width 5cm height0.001pt}}
\vspace*{13pt}
\fcaption{(a) ``Free" evolution of the neutron spin 
under the action of a magnetic field. An emitter $E$ sends a spin-up neutron
through several regions where a magnetic field $B$ is present.
The detector $D_0$ detects a spin-down neutron: No Zeno effect occurs.
(b) The neutron spin is ``monitored"
at every step, by selecting and detecting the spin-down component.
$D_0$ detects a spin-up neutron: The Zeno effect takes place.}
\end{figure}
An incident neutron, travelling along the $y$-direction,
interacts with several identical regions in which there is 
a static magnetic field $B$, oriented along the $x$-direction. 
We neglect unnecessary complications and describe 
the interaction by the Hamiltonian $H= \mu B \sigma_1$
($\mu$ being the (modulus of the) neutron magnetic moment,
and $\sigma_1$ the first Pauli matrix.)
Let the initial neutron state be
$\rho_{0} = \rho_{\uparrow \uparrow} \equiv
\vert \uparrow \rangle\langle\uparrow \vert$, where
$\vert \uparrow\rangle$ and $\vert \downarrow\rangle$
are the spin states of the neutron along the $z$ axis, which 
can be identified with the undecayed and decayed states of the previous 
subsections, respectively.
Assume that there are $N$ regions in which $B$ is present and that the
interaction between the neutron and the magnetic fields has a total duration
$T=Nt$ ($t= \ell/v$, where $\ell$ is the length of each region where 
$B$ is present and $v$ the neutron speed).
It is then straightforward\cite{qze1} to show that 
if $T$ is chosen so as to satisfy the ``matching" condition
$T = (2m+1)\pi/\omega$, where $m$ is an integer 
and $\omega=2 \mu B$, the final state and the probability 
that the neutron spin is found down at time $T$ read respectively
\andy{noZeno}
\barr
\rho (T) & = & \rho_{\downarrow \downarrow},
 \label{eq:noZeno} \\
P_{\downarrow}(T) &= & 1.
 \label{eq:pT}
\earr
The experimental setup in Figure~1(a) is such that if the
system is initially prepared in the up state, it will evolve to
the down state after time $T$ ($\pi$-pulse). 

The situation outlined in Figure~1(b) is very different. The experiment has
been modified by inserting at every step a device able to select and detect the
down component of the neutron spin. This is accomplished by a magnetic mirror
$M$ and a detector $D$. 
The magnetic mirror yields a spectral decomposition\cite{Wigner,mn,np}
by splitting a neutron wave with indefinite spin (a superposed state of up
and down spins) into two branch waves (each of which is in a definite spin 
state along the $z$ axis) and then forwarding the down component to a
detector. The action of the magnetic mirror 
can be compared to the inhomogeneous magnetic field in a
typical Stern-Gerlach experiment. 
It is very important, in connection with the QZE,
to bear in mind that the magnetic mirror does
{\em not} destroy the coherence between the two branch waves:
Indeed, the two branch waves corresponding to different spin states can be
split coherently and brought back to interfere.\cite{strobo}
The global action of $M$ and $D$ can be formally represented by
the operator ${\cal O}\equiv \rho_{\uparrow \uparrow}$.
If the initial $Q$ state and the ``matching" condition for $T=Nt$
are the same as before, the density matrix and the 
probability that the neutron spin is up at time $T=(2m+1)\pi/\omega$ 
read respectively \andy{yesZeno,pZT}
\barr
\rho^{(N)}(T) & = & V_N(T) \rho_0 V_N^\dagger(T) 
  = \left( \cos ^2 {\frac{\omega t}{2}} \right)^N
      \rho_{\uparrow \uparrow}
  = \left( \cos ^2 {\frac{\pi}{2N}} \right)^N
      \rho_{\uparrow \uparrow} ,
\label{eq:yesZeno} \\
P_{\uparrow}^{(N)}(T) & = &
      \left( \cos ^2 {\frac{\pi}{2N}} \right)^N .
\label{eq:pZT}
\earr
This discloses the occurrence of a QZE:
Indeed, $P_\uparrow^{(N)}(T)> P_\uparrow^{(N-1)}(T)$ for $N\geq 2$,
so that the evolution is ``slowed down" as $N$ increases. 
In the limit of infinitely many observations
\andy{NZeno,probfr}
\barr
\rho^{(N)}(T) & \stackrel{N \rightarrow \infty}{\longrightarrow} &
\t{\rho}(T) =  \rho_{\uparrow \uparrow},
\label{eq:NZeno} \\
P_\uparrow^{(N)}(T) & \stackrel{N \rightarrow \infty}{\longrightarrow} &
{\cal P}_\uparrow (T) = 1.
\label{eq:probfr}
\earr
Frequent observations ``freeze" the neutron spin in its initial state, by
inhibiting ($N \geq 2$) and eventually hindering ($N \rightarrow \infty$)
transitions to other states.
Compare Eqs.\ (\ref{eq:NZeno}) and (\ref{eq:probfr})
with (\ref{eq:noZeno}) and (\ref{eq:pT}):
The situation is completely reversed.

It must be observed, however, that it is possible to obtain the 
{\em same} result without making use 
of projection operators, by simply performing a different analysis 
involving {\em only unitary processes}.
Observe first that the preceding analysis involves 
only the $Q$ states. 
If the state of the total (neutron + detectors) system is duely
taken into account, the final total density matrix,
in the channel representation, reads
\andy{big1}
\newfont{\bg}{cmr10 scaled\magstep4}
\newcommand{\bigzerol}{\smash{\hbox{\bg 0}}}
\newcommand{\bigzerou}{\smash{\lower1.7ex\hbox{\bg 0}}}
\beq
\t{\Xi}_{ij} \equiv
\left(
 \begin{array}{ccccc}
   c^{2N}    &               &               &        &\bigzerou \\
             & s^2c^{2N-2}   &               &        &\\
             &               & s^2c^{2N-4}   &        &\\
             &               &               & \ddots &\\
   \bigzerol &               &               &        & s^2 
 \end{array}\right)  , 
   \qquad i,j = 0,1, \ldots ,N ,
\label{eq:big1}
\eeq
where $c = \cos (\pi /2N)$ and $s = \sin (\pi /2N)$.
This corresponds to the case of frequent observations, in which
the neutron route was observed at every step.
The $i=j=0$ component corresponds to detection by $D_0$,
while the $i=j=n \; (n=1, \ldots, N)$ component corresponds to
detection in channel $N-n+1$.
Observe that in the above expression the total density matrix
$\t{\Xi}_{ij}$ has no off-diagonal components as a consequence
of the ``wave function collapse" by measurement.

Remove now $D_1, \cdots,$ $D_N$ in Figure~1(b):
In other words, {\em no} observation of the neutron route
is carried out (except the final one performed by $D_0$).
A straightforward calculation yields the following final density matrix 
for the $Q$ system
\andy{big2}
\beq
\Xi_{ij} \equiv
\left(
  \begin{array}{ccccc}
   c^{2N}      & isc^{2N-1}    & isc^{2N-2}    & \ldots & isc^N     \\
   -isc^{2N-1} & s^2c^{2N-2}   & s^2c^{2N-3} & \cdots & s^2c^{N-1}  \\
   -isc^{2N-2} & s^2c^{2N-3}   & s^2c^{2N-4} & \cdots & s^2c^{N-2}    \\
   \vdots      & \vdots        & \vdots        & \ddots & \vdots      \\
   -isc^{N}    & s^2c^{N-1}    & s^2c^{N-2}      & \cdots & s^2 
  \end{array}
 \right) ,
   \qquad i,j = 0,1, \ldots ,N .
    \label{eq:big2}
\eeq
The two expressions above clearly show that we have the 
{\em same} probability $P_\uparrow^{(N)} \! = 
[\cos^2 (\pi /2N)]^N$ of detecting 
a spin-up neutron at $D_0$ {\em irrespectively} of the presence 
of detectors $D_1, \ldots,$ $D_N$ in Figure~1(b).
It appears therefore that {\em no projection rule is necessary} in this
context. The quantum Zeno effect can be given a purely dynamical explanation.
 
If $N \rightarrow \infty$,
both density matrices tend to the limiting expression 
\andy{big3}
\beq
\Xi^\infty_{ij} \equiv
\left(
  \begin{array}{cccc}
   1 & 0 & 0 & \cdots \\
   0 & 0 & 0 & \cdots \\
   0 & 0 & 0 & \cdots \\
   \vdots  & \vdots & \vdots & \ddots  
  \end{array}
 \right) ,
   \qquad i,j = 0,1, \ldots \ .
    \label{eq:big3}
\eeq
This is the quantum Zeno paradox. It can be obtained  
both by making use of projection operators {\em \`a la} von Neumann
and by means of a purely dynamical process.
The above conclusions can be generalized to an arbitrary quantum system 
undergoing a Zeno-type dynamics, and it can even be shown that 
it is possible to ``mimic" the instantaneous action of a projection operator
by making use of the impulse approximation in quantum
mechanics.\cite{qze1}

It must also be stressed that an idea analogous to the one 
described in this subsection was outlined by Peres,\cite{Peres}
who made use of photons, rather than neutrons.
His proposal inspired an interesting experiment.\cite{inn}

\subsection{The $N \rightarrow \infty$ limit and its physical unrealizability}
\label{sec-Heiss}
\noindent
Let us now discuss the meaning of the $N \rightarrow \infty$ limit
and show that it is unphysical.\cite{qze2}
We start by observing that the condition $\omega T= \omega Nt= (2m+1)\pi$,
which is to be met at every step in Figure~1(b), implies 
(by setting $m=1$ for simplicity and without loss of generality)
\andy{condBl}
\beq
B \ell = \frac{\pi \hbar v}{2 \mu N} = O(N^{-1}) ,
    \label{eq:condBl}
\eeq
where all quantities were defined before Eq.\ (\ref{eq:noZeno}). Obviously, 
as $N$ increases in the above equation, the {\em practical} realization of the
experiment becomes increasingly difficult.
But close scrutiny of
Eq.\ (\ref{eq:pZT}) shows that $P_{\uparrow}^{(N)}(T)$
cannot tend to 1, even {\em in principle}, 
in the $N \rightarrow \infty$ limit, because of the uncertainty relations.
Indeed, let $\phi$ 
be the argument of the cosine in Eq.\ (\ref{eq:pZT}):
\andy{argy}
\beq
\phi \equiv \frac{\mu B \ell}{\hbar v} = \frac{\pi}{2N} .
\label{eq:argy}
\eeq
Mathematically, the above quantity is of order $N^{-1}$. 
On the other hand, from a {\em physical} 
point of view, it is impossible to avoid uncertainties in the neutron position
$\triangle x$ and speed $\triangle v$.
As a consequence, $\phi$ cannot vanish, because it is lower bounded as follows
\andy{lowb}
\beq
\phi \simeq \phi_0 = \frac{\mu B \ell}{\hbar v_0} 
    > \frac{\mu B \triangle x}{\hbar v_0} 
    > \frac{\mu B }{ 2M v_0 \triangle v},
\label{eq:lowb}
\eeq
where $M$ is the neutron mass and we assumed that the size $\ell$ of the 
interaction region (where the neutron
spin undergoes a rotation under the action of the magnetic field) is 
larger than the longitudinal spread $\triangle x$
of the neutron wave packet.
An accurate analysis\cite{qze2} shows that the same bound holds
in the opposite situation $(\ell < \triangle x)$ as well.

By defining the magnetic energy gap $\triangle E_m = 2 \mu B$
and the kinetic energy spread of the neutron beam 
$\triangle E_k = \triangle (M v^2/2) |_{v=v_0}$, the above inequality reads
\andy{lowbb}
\beq
\phi > \frac{1}{4}\frac{\triangle E_m}{\triangle E_k}.
\label{eq:lowbb}
\eeq
It is now straightforward to obtain an expression for the value of the 
probability that a spin-up neutron is observed at $D_0$ when $N$ is large:
\andy{pnb}
\beq
P_{\uparrow}^{(N)}(T) \simeq 
      \left( \cos \phi_0 \right)^{2N} \simeq 
      \left( 1 - \frac{1}{2} \phi_0^2 \right)^{2N} 
   \simeq \left[ 1 - \frac{1}{32}
        \left( \frac{\triangle E_m}{\triangle E_k} \right)^2 \right]^{2N} .
\label{eq:pnb}
\eeq
Notice that not only the above quantity does {\em not} tend to 1, but it 
{\em vanishes} in the $N \rightarrow \infty$ limit.
In other words, in the experiment
outlined in Figure~1(b), {\em no} spin-up neutron would
be observed at $D_0$ in the $N \rightarrow \infty$ limit!
 
What is it reasonable to expect, in practice?
Even though the analysis of the previous subsections does not 
take into account the limits imposed by the uncertainty principle,
it must be considered that, in practice, $N$ cannot be made arbitrarily 
large. In order to evaluate how big $N$ can be in order that  
the QZE be observable in the above experiment, set
$P_{\uparrow}^{(N)}(T) \sim 1/2$.
We get
\andy{Nbo}
\beq
N \simeq \frac{64 \ln 2}{(\triangle E_m/\triangle E_k)^2} \sim 10^4,
\label{eq:Nbo}
\eeq
where we assumed 
reasonable values for the energies of a thermal neutron.
In conclusion, $N$ turns out to be large enough in order that the QZE be
experimentally observable, at least up to a certain approximation.
 
Criticisms against the physical meaning and realizability of the $N 
\rightarrow 
\infty$ limit were put forward some years ago by Ghirardi 
et al.\cite{critico} Although different from ours, these criticisms were based
on the time-energy uncertainty relations. Our argument, outlined for neutrons,
holds true in general, and even in the recent experiment
performed with photons in Innsbruck.\cite{inn} An exhaustive 
analysis is given in Ref.\ \citelow{qze2}.
 
There are in fact other strong arguments against the 
$N \rightarrow \infty$ limit, from a physical point of view.
For instance, the above calculation considers only the time spent by the 
neutron in the magnetic field $B$. In practice, however, one cannot 
neglect the time elapsed during the interaction between the neutron and
a magnetic mirror M, which is of the order of $10^{-6}-10^{-7}$ s.

The above discussion should have made it clear that although the short-time 
behavior is essentially governed by an evolution of the Gaussian type,
the limit of continuous observation ($N \rightarrow \infty$)
should never be lightheartedly taken for granted.
It is safer and much more reasonable to start one's considerations
from formulas like (\ref{eq:spdy}), where both $T$ and $N$ are finite.

On the other hand, we shall soon see that there are physical 
situations in which the temporal behavior
of a quantum system can be considered approximately exponential
(the mathematical and physical conditions of validity of such 
an assumption will be thoroughly discussed in the following section).
In such a case, the survival probability at time $T$ after $N$ measurements
would read
\andy{spde}
\beq
P_{{\rm E}}^{(N)}(t) \simeq \left( 1-\frac{1}{\tau_{\rm E}}
\left(\frac{T}{N} \right) \right)^N \simeq
e^{-T/\tau_{\rm E}}.
\label{eq:spde}
\eeq
where the subscript $_{\rm E}$ stands for ``exponential", and 
$\tau_{\rm E}$ is the characteristic time of the exponential law.
Notice that, unlike in (\ref{eq:spdy}),
in this case the final result {\em does not strongly depend on $N$}.

If there exist finite values of $T$ and $N$ such that 
\andy{compp}
\beq
\frac{T}{\tau_{\rm E}} \simeq \frac{T^2}{\tau_{\rm G}^2N}
\quad \Longleftrightarrow \quad
\frac{T}{N\tau_{\rm G}} \simeq \frac{\tau_{\rm G}}{\tau_{\rm E}},
\label{eq:compp}
\eeq
then the exponential and the Gaussian regions are comparable with each other.
Which temporal behavior is actually to be observed depends therefore 
on the relative
magnitude of the parameters $\tau_{\rm G}$ and $\tau_{\rm E}$.
For this reason 
we may not completely exclude the possibility of observing a sort of quantum
Zeno effect within the framework of the exponential decay.
An analogous point was raised by Schulman, Ranfagni and Mugnai,\cite{Schulman}
within the context of the WKB approximation.

It should be emphasized that it is very difficult to obtain general 
estimates of the two characteristic times mentioned above.
For example, Khalfin, as well as other authors, even considered 
and critically discussed 
the possibility that the proton decay has never been observed because
its Gaussian characteristic time 
$\tau_{\rm G}^{\rm prot}$ might be longer than 
the lifetime of the Universe.\cite{protdec}

\setcounter{equation}{0}
\section{Quantum behavior at longer times}
\label{sec-longt}
\noindent
In the first part of the preceding section, we have seen that the 
temporal behavior of the survival probability at short times is well
described by a Gaussian form.  Our main task in this section
is to investigate the temporal behavior of quantum systems at longer 
times and to understand 
the status of the familiar exponential law in quantum theory. 

Since an articulate and complete analysis of the temporal behavior
of quantum mechanical systems, especially at very long times, is a 
rather complex issue, we shall organize our discussion in the following way.
First, in Sec.\ \ref{sec-longt}.\ref{sec-naive}, 
we consider a particular interaction Hamiltonian and discuss
the temporal behavior both at short and long times.
Then, in Sec.\ \ref{sec-longt}.\ref{sec-generala}, we reconsider 
our conclusions in the most general case and discuss 
which problems arise in the very long time region as a consequence 
of the analytic properties of the survival amplitude. 
Finally, for the sake of clarity,
in Sec.\ \ref{sec-longt}.\ref{sec-nonexp}, we return to the particular
case and thoroughly discuss the very long time behavior.

Suppose that the total Hamiltonian is decomposed into two parts, 
$H=H_{0}+H'$, where $H_0$ and $H'$ are the unperturbed and
the interaction Hamiltonians, respectively. 
Let $\vert n\rangle$ be an eigenstate of the former belonging to 
the eigenvalue 
$E_n$
\andy{H0}
\beq
  H_0\vert n\rangle=E_n\vert n\rangle,\qquad
  1=\sum_n\vert n\rangle\langle n\vert.
\label{eq:H0}
\eeq
Even though we use the discrete notation for notational simplicity, 
it is to be understood that the possibility of continuous spectrum 
is included as well.
At $t=0$, the interaction is turned on and the system starts to evolve 
from an initial state, say $\vert \alpha \rangle$, into the other states.
The initial state is not an eigenstate 
of $H$, and is prepared as an appropriate wave packet, that can be 
represented by a superposition of $\vert n\rangle$'s. 
For the sake of simplicity, we shall denote the initial state by
an eigenstate of $H_{0}$, say $\vert a\rangle$.
The interaction Hamiltonian $H'$ is chosen so that the condition
\andy{nH'n} 
\beq
  \langle n\vert H'\vert n\rangle=0
\label{eq:nH'n}
\eeq
is satisfied $\forall n$.
This ensures that the Lippmann-Schwinger equation
\andy{LS}
\beq
\vert\psi_a\rangle=\vert a\rangle+{1\over E_a-H_0}H'\vert\psi_a\rangle
\label{eq:LS}
\eeq
can be solved.
Notice that the condition (\ref{eq:nH'n}) corresponds to mass 
renormalization in field theories. As already mentioned,
we shall start by discussing the temporal behavior of 
the survival probability amplitude in a simple case.

\subsection{Derivation of temporal behaviors: Standard textbook method}
\label{sec-naive}
\noindent
Here we shall choose a particular interaction Hamiltonian $H'$ 
subject to the condition that it
has nonvanishing matrix elements only between 
the initial state $\vert a\rangle$ and the 
other states $\vert n\rangle$ ($n\not=a$):
\andy{matrixelement} 
\beq
  \langle n\vert H'\vert n'\rangle=0,\quad
  \langle a\vert H'\vert n\rangle\not=0,\quad
  (n,n'\not=a).
\label{eq:matrixelement}
\eeq
We shall see later that the first relation can be considered as a stronger 
version of the so-called random phase approximation (see (\ref{eq:nH'n'}) 
below).
Observe that these conditions are peculiar to some solvable models.
Historically, the Lee model was based on a similar idea.\cite{Lee}
Nevertheless, the above can still be considered as a rather general 
form of the interaction Hamiltonian, as can be seen in some standard 
textbook.\cite{Messiah1959} 

     We are interested in the temporal behavior of the survival probability 
amplitude $\langle a\vert U_I(t)\vert a\rangle$.
The evolution operator in the interaction picture $U_I(t)$ satisfies the 
Schr\"odinger equation
\andy{Udot}
\beq
i{d\over dt}U_I(t)=H'_I(t)U_I(t),\qquad H'_I(t)=e^{iH_0t}H'e^{-iH_0t},
\label{eq:Udot}
\eeq
where $H'_I(t)$ is the interaction Hamiltonian in the interaction picture.
The solution $U_I(t)$ can be obtained iteratively from 
the recursion relation
\andy{Urecursion}
\barr 
  U_I(t)&=&1-i\int_0^tdt'H_I(t')U_I(t') \nonumber \\
        &=&1-i\int_0^tdt'H'_I(t')
            +(-i)^2\int_0^tdt_1\int_0^{t_1}dt_2H'_I(t_1)H'_I(t_2)U_I(t_2),
\label{eq:Urecursion} 
\earr 
under the initial condition $U_I(0)=1$.
Notice that the conditions (\ref{eq:nH'n}) and (\ref{eq:matrixelement}) 
for the interaction Hamiltonian imply
\andy{H'H'U}
\beq
  \langle a\vert H'_I(t_1)H'_I(t_2)U_I(t_2)\vert a\rangle
  =\langle a\vert H'_I(t_1)H'_I(t_2)\vert a\rangle
   \langle a\vert U_I(t_2)\vert a\rangle.
\label{eq:H'H'U}
\eeq
This relation enables us to obtain a closed equation for the diagonal matrix 
element of the 
evolution operator $\langle a\vert U_I(t)\vert a\rangle$ (survival amplitude) 
\andy{aUa}
\beq
  \langle a\vert U_I(t)\vert a\rangle
  =1-\int_0^tdt_1\int_0^{t_1}dt_2e^{iE_a(t_1-t_2)}f(t_1-t_2)
   \langle a\vert U_I(t_2)\vert a\rangle,
\label{eq:aUa}
\eeq
or an integro-differential equation
\andy{aUadot}
\beq
  {d\over dt}\langle a\vert U_I(t)\vert a\rangle
  =-\int_0^tdt_1e^{iE_a(t-t_1)}f(t-t_1)\langle a\vert U_I(t_1)\vert a\rangle
\label{eq:aUadot}
\eeq
with the initial condition $\langle a\vert U_I(0)\vert a\rangle=1$.
Here the function $f(t)$ is defined as
\andy{f}
\beq
  f(t)=\langle a\vert H'e^{-iH_0t}H'\vert a\rangle\ .
\label{eq:f}
\eeq
Note that (\ref{eq:aUa}) is a series of repeated convolutions of 
$e^{iE_{a}(t_{1}-t_{2})}f(t_{1}-t_{2})$, and 
all other higher-order terms disappear, due
to the particular conditions (\ref{eq:matrixelement}).
From this point of view, it is convenient to
reduce (\ref{eq:aUa}) or 
(\ref{eq:aUadot}) to an algebraic equation by means of a Laplace 
transform, which is quite appropriate 
in order to incorporate the initial condition.
The solution can now be written as an inverse Laplace transform
\andy{Laplace}
\beq
  \langle a\vert U_I(t)\vert a\rangle={1\over2\pi i}
  \int_{-i\infty+\epsilon}^{i\infty+\epsilon}{e^s\over g(s,t)}ds ,
\label{eq:Laplace}
\eeq
where the function $g$ is given by
\andy{g}
\beq
  g(s,t)=s+t\int_0^\infty e^{-{s\over t}u}e^{iE_au}f(u)du.
\label{eq:g}
\eeq
(The second term in $g(s,t)$ will turn out to correspond to the self-energy 
part $\Sigma_a(E)$ in the general case.
See the next subsection.)

     Having obtained the solution 
in a closed form, we can discuss its temporal behavior and extract its
explicit time dependence both at short and long times.
For small $t$, the function $g$ is expanded around $t=0$
\andy{smallt}
\barr
  g(s,t)&=&s+t^2\int_0^\infty e^{-su}e^{iE_atu}f(tu)du\nonumber\\
        &=&s+t^2\int_0^\infty e^{-su}
        \left(f(0)+{d\over dx}\left.
               \left[ e^{iE_ax}f(x)\right]\!\right|_{x=0}\!tu 
                                         +\cdots\right)du\nonumber\\
        &\simeq&s+{f(0)\over s}t^2,
\label{eq:smallt}
\earr
which determines the behavior at small $t$ 
\andy{aUasmallt}
\barr
  \langle a\vert U_I(t)\vert a\rangle
  &\simeq&{1\over2\pi i}\int_{-i\infty+\epsilon}^{i\infty+\epsilon}
    {se^s\over s^2+f(0)t^2}ds=\cos\sqrt{f(0)}t                \nonumber\\
  &\simeq&1-{f(0)\over2}t^2 \simeq e^{-{f(0)\over2}t^2}  .
\label{eq:aUasmallt}
\earr
Notice that $f(0)=\langle a\vert H'^2\vert a\rangle$ is positive definite.
The system exhibits the Gaussian behavior for small $t$, in accordance with the general argument stated in Sec.~\ref{sec-stbeh}.
[Specialize Eq.\ (\ref{eq:naie}) in Sec.~\ref{sec-stbeh} to the case in which
$|a\rangle$ is an eigenstate of $H_0$, under the condition
(\ref{eq:nH'n}).]

     On the other hand, for large $t$, we can see that the amplitude 
$\langle a\vert U_I(t)\vert a\rangle$ behaves quite differently.
In this case, the function $g$ is expanded as a power series of $1/t$
\andy{larget}
\barr
  g(s,t)&=&s+t\int_0^\infty e^{-{s\over t}u}e^{iE_au}f(u)du\nonumber\\
        &=&s+t\int_0^\infty 
           \left(1-{s\over t}u+\cdots\right)e^{iE_au}f(u)du\nonumber\\
        &\simeq&s\left(1-\int_0^\infty ue^{iE_au}f(u)du\right)
              +t\int_0^\infty e^{iE_au}f(u)du.
\label{eq:larget}
\earr
Therefore we obtain the following exponential behavior at longer times
\andy{aUalarget}
\beq
  \langle a\vert U_I(t)\vert a\rangle\sim e^{-\Lambda t},
\label{eq:aUalarget}
\eeq
as long as the exponent $\Lambda$, given by
\andy{exponent}
\beq
  \Lambda={\int_0^\infty e^{iE_au}f(u)du
           \over
           1-\int_0^\infty ue^{iE_au}f(u)du},
\label{eq:exponent}
\eeq
exists.
The derivation explained here is similar to that given by 
Messiah,\cite{Messiah1959} even though we endeavoured to bridge the gap 
with the general theory to be sketched in the following subsection.
Notice that the above form (\ref{eq:aUalarget}) does {\em not} necessarily 
mean that the survival probability {\em decays exponentially}. In order to
obtain such a behavior we have to show that the real part of the exponent 
$\Lambda$ is positive.

     The positivity of the real part of the exponent $\Lambda$ 
cannot always be shown, as can be seen in some simple solvable models, where 
only the oscillating behavior appears at all times.
Consider, for example, the Hamiltonian $H=H_0+H'=\mu B_0 \sigma_3 +
\mu B \sigma_1$, where $H_0=\mu B_0 \sigma_3$ is regarded as 
the free part and $H'=\mu B \sigma_1$ as the interaction (a similar 
example was considered in the context of the quantum Zeno effect,
in the paper by Inagaki et al.\cite{Prigogine}).
If we start from the initial state $|a \rangle = | \uparrow \rangle$,
the above analysis yields 
(see (\ref{eq:f}) for the definition of $f$)
$f(t) = (\mu B)^2e^{i\mu B_0t}$ and a purely
imaginary $\Lambda=i \mu B_0 B^2 / (2B_0^2+B^2)$.
As was to be expected, Eq.\ (\ref{eq:aUalarget}) yields an
oscillatory behavior at all times and no exponential
law can be seen. This is ascribable to the
finiteness of the system considered: We cannot expect any exponential behavior
if the number of degrees of freedom of the system is finite. As a matter 
of fact, the numerator appearing in $\Lambda$ in (\ref{eq:exponent}) is 
purely imaginary unless the initial state $\vert a\rangle$ is continuosly 
degenerated w.r.t.\ energy.
We have 
\andy{numerator}
\barr
  \int_0^\infty e^{iE_au}f(u)du
  &=&\lim_{\epsilon\to+0}
     \langle a\vert H'{i\over E_a-H_0+i\epsilon}H'\vert a\rangle\nonumber\\
  &=&\langle a\vert H'\mbox{P}{i\over E_a-H_0}H'\vert a\rangle
     +\pi\langle a\vert H'\delta(E_a-H_0)H'\vert a\rangle\nonumber\\
  &=&i\sum_{E_n\not=E_a}{\vert\langle n\vert H'\vert a\rangle\vert^2
                             \over E_a-E_n}
     +\pi\sum_{E_n=E_a,n\not=a}\vert\langle n\vert H'\vert a\rangle\vert^2.
\label{eq:numerator}
\earr
This expression implies that if $\vert a\rangle$ is not degenerated the 
last term gives no contribution, which makes the above quantity and therefore 
the exponent $\Lambda$ purely imaginary.
The degeneracy of the initial state is necessary for the survival 
probability to decay exponentially.
Remember that we are mainly concerned with a 
particle system coupled with fields in free space or with an interacting 
system made up of a particle and a many-body system with a huge number of 
degrees of 
freedom, put in a finite but very large box, as was mentioned in the 
introductory section. 
For such systems, we naturally expect the continuous energy spectrum 
to be always degenerated.

It is worth noting that the above expression is a realization of the 
famous Fermi Golden Rule,\cite{Fermigold} according to which the decay 
rate is given by
\andy{goldenrule}
\beq
  \Gamma=2\pi\sum_{E_n=E_a,n\not=a}\vert\langle n\vert H'\vert a\rangle\vert^2
\label{eq:goldenrule}
\eeq
by first-order perturbation theory.
We understand that twice the real part of $\Lambda$, standing for our decay 
rate, coincides with the above $\Gamma$ in the weak-interaction limit
(so that the denominator in (\ref{eq:exponent}) is approximately equal to one).
Needless to say, this formula can be improved by including higher-order
perturbation terms in a well-known way.

     Although the conclusion we have obtained seems reasonable and 
its derivation sound and standard,\cite{Messiah1959} it is known that 
at very long times 
quantum systems never show the exponential behavior, a fact in contradiction 
with the above result as well as with our naive expectation.
In fact, the unattainability of the exponential law
at very long times is a mathematically unavoidable consequence. 
In order to see this,
recall that the evolution operator in the interaction picture $U_I(t)$ 
is expressed as
\andy{Ufactored}
\beq
  U_I(t)=e^{iH_0t}e^{-iHt},
\label{eq:Ufactored}
\eeq
in terms of the total Hamiltonian $H$. By 
introducing a complete orthonormal set of eigenstates of the latter, 
we easily arrive at the 
following form for the survival probability amplitude
\andy{aUaFT}
\beq
  \langle a\vert U_I(t)\vert a\rangle=e^{iE_at}\int\omega_a(E)e^{-iEt}dE,
\label{eq:aUaFT}
\eeq
where $\omega_a(E)$ is the energy density of the initial state
\andy{omega}
\beq
  \omega_a(E)\equiv\sum_{E_n=E}\vert\langle n\vert a\rangle\vert^2.
\label{eq:omega}
\eeq
Here the summation is taken over all the quantum numbers,
except energy, that are necessary for the 
specification of a complete orthonormal set.
We see that the survival probability amplitude is the Fourier 
transform of the energy density $\omega_a(E)$, a fact first pointed out by 
Fock and Krylov.\cite{FockKrylov}\ 
Now let us assume, on physical grounds, that the spectrum of the total 
Hamiltonian is bounded from below so that the vacuum state is stable:
There exists a certain finite energy $E_g$ below which the function 
$\omega_a(E)$ vanishes.
Khalfin showed\cite{Khalfin1957+58} that this 
condition on $\omega_a$ requires that its Fourier transform (the survival 
probability amplitude) satisfies the inequality
\andy{PaleyWiener}
\beq
  \int_{-\infty}^\infty
  {\bigl\vert\ln\vert\langle a\vert U_I(t)\vert a\rangle\vert\bigr\vert
        \over1+t^2}dt<\infty,
\label{eq:PaleyWiener}
\eeq
as a consequence of the fundamental Paley-Wiener Theorem.%
\cite{PaleyWiener}\ 
The inequality (\ref{eq:PaleyWiener}) implies that the survival 
probability does not decay exponentially:
The decay process proceeds more slowly than exponentially at large 
times.
Notice that the only assumption made in the above argument is the 
existence of a finite $E_g$:
Its very value is irrelevant.
Therefore the conclusion is quite general.\fnm{d}\fnt{d}{A full 
account of this theorem, related topics and further references 
are given in Ref.~\citelow{Exner}.}

     At this point, we may feel somewhat puzzled with the above results, 
especially with that at long times, because we are 
familiar with the exponential decay law in classical theory and naturally 
expect the same form to be valid also in quantum theory.
Indeed, the exponential law for decay processes has been well confirmed 
experimentally:
No deviation from it at long times has ever been observed.
A list of experimental results, examining the exponential law, can be found 
in Ref.~\citelow{YY93b}.
Then how can we reconcile quantum theory with the exponential law?

  The above-mentioned theorem (\ref{eq:PaleyWiener}) requires that the 
amplitude decays more slowly than exponentially in the large-time limit 
$t\to\infty$.
Therefore we may expect it to decay exponentially at intermediate times, 
even if not at very long times.
At any rate, something must have been overlooked in our derivation of 
(\ref{eq:aUalarget}).
In the next subsection, we shall reanalyze the temporal behavior in full 
generality and shall focus our attention on the 
analytic properties of the amplitude: Indeed, in the above analysis, we have 
assumed that the order of the integrations over $s$ and $u$ and the 
$t \rightarrow 0$ and $t \rightarrow \infty$ limits can be interchanged,
which is licit only when the integrals converge uniformly and the initial 
state is a well-behaved wave packet. This is by no means trivial,
and is where subtleties might come into play.

\subsection{General arguments}
\label{sec-generala}
\noindent
Let us analyze in full generality the temporal behavior of the survival 
probability amplitude $\langle a\vert U_I(t)\vert a\rangle$.
The evolution operator in the interaction picture $U_I(t)$ is 
related to its counterpart $U(t)$ in the Heisenberg picture and to
the $S$-matrix in the usual way:
\andy{UUS}
\beq
U(t)=e^{-iHt}=e^{-iH_0t}U_I(t)
\stackrel{t\rightarrow \infty}{\longrightarrow} e^{-iH_0t}S\ ,
\label{eq:UUS}
\eeq
where $H$ and $H_{0}$ stand for the total and the free Hamiltonians, 
respectively.
It is important to notice that the unitary operator $U(t)$ 
is expressed as a Fourier transform
\andy{FU}
\beq
U(t)=\frac{i}{2\pi}\int_Ce^{-iEt}G(E)dE;\qquad 
G(E)=\frac{1}{E-H}\ ,
\label{eq:FU}
\eeq
and a similar relation holds 
for its free counterpart $U_{0}(t)$, with $G_{0}(E)$.

Note that the analytic structure of $G(E)$ [$G_{0}(E)$] in the 
complex $E$-plane is symmetric w.r.t.\ the real $E$ axis, as a 
reflection of the time-reversal invariance of the whole process.
Because of the relation 
$\vert e^{-iEt}\vert=\exp[\vert E\vert t\sin(\arg E)]$, we know 
that $U(t)$ [$U_{0}(t)$] vanishes for $t<0$ if the integration contour $C$, 
running from $-\infty$ to $\infty$ along the 
real $E$ axis, lies a little above all the singularities of $G(E)$
[$G_{0}(E)$].  For the initial-value 
problem (Cauchy problem), therefore, we have to choose 
the integration contour $C$ in the way shown in 
Figure~\ref{fig:singGdis}.

\begin{figure}[htbp]
\vspace*{13pt}
\centerline{\vbox{\hrule width 5cm height0.001pt}}
\vspace*{1.4truein}             
\centerline{\vbox{\hrule width 5cm height0.001pt}}
\vspace*{13pt}
\fcaption{Singularities of $G(E)$ or $G_{0}(E)$ for the case of discrete 
spectrum and the integration contour $C$ for the initial-value problem.}
\label{fig:singGdis}
\end{figure}

Let us start our discussion from the case of a system composed of 
a finite number of particles (say, $N$) put in a finite box (of volume $V$).
Then $H$  $[H_{0}]$ has a 
discrete spectrum and, correspondingly, the singularities of $G(E)$ 
$[G_{0}(E)]$ appear as a series of simple poles, running from the 
lowest-energy point (i.e.\ the lower bound of the spectrum) 
to infinity on the real $E$ axis. We denote the 
lowest-energy point in both cases by $E_g$, for simplicity. 
See Figure~\ref{fig:singGdis}.

We can now start the perturbation theory on the
basis of the following expansion
\andy{Gexp}
\barr
G(E)&=&\frac{1}{E-H_0}
       +\frac{1}{E-H_0}H'\frac{1}{E-H_0}\nonumber\\
    & &+\frac{1}{E-H_0}H'\frac{1}{E-H_0}H'\frac{1}{E-H_0}\nonumber \\
    & &+\frac{1}{E-H_0}H'\frac{1}{E-H_0}H'\frac{1}{E-H_0}H'\frac{1}{E-H_0}
                                                                \nonumber \\
    & &+\frac{1}{E-H_0}H'\frac{1}{E-H_0}H'\frac{1}{E-H_0}H'\frac{1}{E-H_0}H'
        \frac{1}{E-H_0}\nonumber \\
    & &+\cdots.
\label{eq:Gexp}
\earr
The survival probability amplitude (in the Heisenberg picture) is given 
by\fnm{e}\fnt{e}
{The quantity $G_a(E)$ is essentially the same as the energy density 
$\omega_a(E)$ in (\ref{eq:omega}), though the meaning of $E$ is different:
The latter $E$ stands for the 
spectrum of the total Hamiltonian $H$ while the former $E$ is just an 
integration variable running from $-\infty$ to $\infty$ in the Fourier 
transform, irrespectively of the spectrum of $H$.}\  
\andy{Ga}
\begin{equation}
\langle a |U(t)| a \rangle =\frac{i}{2\pi}\int_{C} e^{-iEt} G_{a}(E) dE,
\qquad
G_{a}(E)\equiv \langle a |G(E)| a \rangle.
\label{eq:Ga}
\end{equation}

We introduce now the plausible assumption that the phases of the off-diagonal 
matrix elements of the interaction Hamiltonian $H'$ are randomized in the 
infinite $N$ and $V$ limit (keeping $N/V$ finite). That is, 
\andy{nH'n'}
\beq
\langle n\vert H'\vert n'\rangle\quad\mbox{have random phases for}\ n\neq n'.
\label{eq:nH'n'}
\eeq

Note that the discrete energy spectrum becomes continuous in this limit. 
Consequently, the above-mentioned simple discrete poles of 
$G_{0}(E)$ get closer and closer in this limit, and finally merge into a 
continuous line, distributing over $[E_g,\infty)$ along the real axis, 
where $E_g$ is the lowest-energy point of the free Hamiltonian. 
In this case, there is a branch point at $E_g$ and a branch cut on 
$[E_g, \infty)$ for $G(E)$ and, in particular, for the self-energy part 
$\Sigma_{a}(E)$ to be introduced shortly.
See Figure~\ref{fig:singGcon}.

\begin{figure}[htbp]
\vspace*{13pt}
\centerline{\vbox{\hrule width 5cm height0.001pt}}
\vspace*{1.4truein}             
\centerline{\vbox{\hrule width 5cm height0.001pt}}
\vspace*{13pt}
\fcaption{Singularities of $G_{a}(E)$ for the continuous spectrum 
and the integration contour for the initial-value problem. 
The branch cut reflects the continuous spectrum of the free Hamiltonian
and runs along the real $E$ axis.}
\label{fig:singGcon}\andy{fig:singGcon}
\end{figure}

Under the above random phase assumption, it is easy to show that $G_{a}(E)$, 
which is nothing but the Fourier transform of the survival amplitude, is 
written as
\andy{aGa}
\barr
G_a(E)&=&\frac{1}{E-E_a}
         +\left(\frac{1}{E-E_a}\right)^2
          \langle a\vert(H'+H'\frac{1}{E-H_0}H'+\cdots)\vert a\rangle
                                                             \nonumber\\
      &=&\frac{1}{E-E_a}\left[1+\left(\frac{\Sigma_a(E)}{E-E_a}\right)
         +\left(\frac{\Sigma_a(E)}{E-E_a}\right)^2+\cdots\right]\nonumber\\
      &=&\frac{1}{E-E_a-\Sigma_a(E)}.
\label{eq:aGa}
\earr
The ``self-energy part" $\Sigma_a(E)$ is composed of all the even-order 
contributions
\andy{SigE}
\beq
\Sigma_a(E)=\Sigma_a^{(2)}(E)+\Sigma_a^{(4)}(E)+\cdots,
\label{eq:SigE}
\eeq
where
\andy{Sig2,Sig4}
\barr
\Sigma_a^{(2)}(E)
 &=&\langle a\vert H'\frac{1}{E-H_0}H'\vert a\rangle
  =\sum_{n\neq a}\frac{\vert\langle a\vert H'\vert n\rangle\vert^2}{E-E_n},\\ 
\label{eq:Sig2}
\Sigma_a^{(4)}(E)
 &=&\langle a\vert H'\frac{1}{E-H_0}H'\frac{1}{E-H_0}H'\frac{1}{E-H_0}H'
    \vert a\rangle-{1\over E-E_a}\left[\Sigma_a^{(2)}(E)\right]^2\nonumber\\
 &=&\sum_{n\neq a}\sum_{n'\neq a,n}
    \frac{\vert\langle a\vert H'\vert n'\rangle\vert^2\cdot
          \vert\langle n'\vert H'\vert n\rangle\vert^2}{(E-E_n)(E-E_{n'})^2},
\label{eq:Sig4}
\earr
and so on.
Observe that these quantities correspond to the ``proper" self-energy parts 
in field theory (the ``improper" parts are subtracted at every step of the
expansion).
In general, we obtain
\andy{Sig}
\beq
\Sigma_a(E)
 =\sum_{n\not=a}{1\over E-E_n}
  \left[\vert\langle a\vert H'\vert n\rangle\vert^2
        +\sum_{n'\not=a,n}{\vert\langle a\vert H'\vert n'\rangle\vert^2\cdot
         \vert\langle n'\vert H'\vert n\rangle\vert^2\over(E-E_{n'})^2}
        +\cdots\right].
\label{eq:Sig}
\eeq
In the infinite $N$ and $V$ limit, of course, the summation in (\ref{eq:Sig})
becomes the following integral:
\andy{Sigint}
\beq
\Sigma_a(E)
 =\int_{E_g}^{\infty}dE' \rho_{a}(E') \frac{1}{E-E'}K_{a}(E')\ ,
\label{eq:Sigint}
\eeq
where $\rho_{a}(E')$ stands for the state number density and $K_{a}(E')$ 
for the quantity inside the square brackets $[\cdots]$ in (\ref{eq:Sig}).
This expression implies that $G_{a}(E)$ has a branch point at $E_g$ 
and the corresponding branch cut on the real axis, and has no other 
singularities on the first Riemannian sheet, while on the second Riemannian 
sheet possible singularities can appear.
See Figure~\ref{fig:singGcon}.
This procedure, based on the above-mentioned random phase approximation,
is the same as that originally introduced in the theory of nuclear 
reactions.\cite{BlochNamiki}\ It is also very close to the 
{\it diagonal singularity} formulated by van Hove,\cite{vanHove} 
although we have not yet used his limit concerning 
the time-scale transformation.

     The survival probability amplitude (\ref{eq:aGa}), 
with the above $\Sigma_{a}(E)$, has been derived 
under the random phase assumption (\ref{eq:nH'n'}) for 
many-body systems having a huge but finite number of degrees of freedom. 
However, 
as is well known, it is an exact relation in field-theoretical cases 
where we have a particle system coupled to fields in free space or 
two (or more) fields coupled to each other. Of course, in such cases 
the number of degrees of freedom is infinite. This is the reason why 
the quantity $\Sigma_a(E)$ has been called the ``self-energy part".
Investigation of the analytic properties of the amplitude or the relevant 
$S$-matrix elements on the basis of quantum field theory can be traced back to 
the damping theory initiated by Heitler:\cite{Heitler}\ See the review 
paper given by Arnous and Zienau.\cite{ArnousZienau}\ 
Therefore, we can discuss 
the temporal evolution of the survival probability amplitude, in the wider
framework of field theory and many-body problems, on the basis of the 
expression we have just obtained.

It is worth pointing out that the familiar Breit-Wigner\cite{BW} 
(or Weisskopf-Wigner\cite{WW}) form does neither match nor 
follow from the above analysis, since the Breit-Wigner spectrum, 
which is just assumed on a 
phenomenological basis, extends from $-\infty$ to $\infty$.
Also, notice that if we put $E=E_a$ in $\Sigma_a(E)$ (which is in general
a complex number), 
Eq.\ (\ref{eq:aGa}) yields the Weisskopf-Wigner formula\cite{WW} of 
the exponential decay. Of course, this cannot be justified even in the 
weak-coupling limit, since the exponent should be given by the 
simple pole of $G(E)$, say $\bar E$, satisfying
\andy{Ebar}
\beq
\bar E=E_a-\Sigma_a(\bar E),
\label{eq:Ebar}
\eeq
which is clearly different from the former, $\bar E\not=E_a-\Sigma_a(E_a)$.

In order to observe the temporal behavior of the survival probability 
amplitude at longer times, 
the original contour $C$, running along the real $E$ axis from $-\infty$ to 
$\infty$, is now deformed into a new contour.
Recalling again that $G_a(E)$ has a branch cut along the real $E$ axis from 
$E_g$ to $\infty$ and a simple pole on the second Riemannian sheet,
we understand that the $(E_g,\infty)$ portion of $C$ is equivalent 
to the sum of a path running just below the real $E$ axis on the second 
Riemannian sheet from $E_g$ to $-\infty$ and a circle turning  clockwise
around the simple pole. The
new contour, equivalent to the original one, is thus composed of the path 
$C'$ and the circle on the second sheet, as shown in Figure~\ref{fig:scRs}.
\begin{figure}[htbp]
\vspace*{13pt}
\centerline{\vbox{\hrule width 5cm height0.001pt}}
\vspace*{1.4truein}             
\centerline{\vbox{\hrule width 5cm height0.001pt}}
\vspace*{13pt}
\fcaption{The deformed integration contour, composed of a path $C'$ and 
a circle on the second Riemannian sheet.
The path $C'$ starts from $-\infty$, runs just above the real axis 
on the first sheet, goes down to the second sheet at $E_g$ and runs back to 
$-\infty$ just below the real axis.
The circle encloses the simple pole clockwise on the second sheet.}
\label{fig:scRs}\andy{fig:scRs}
\end{figure}
We therefore obtain from Eqs.\ (\ref{eq:Ga}), (\ref{eq:aGa}) and 
(\ref{eq:Ebar}) 
\andy{scRs}
\begin{eqnarray}
\langle a\vert U(t)\vert a\rangle
&=&\frac{i}{2\pi}\int_{C} e^{-iEt}G_{a}(E) dE \nonumber \\
&=&Ze^{-i\bar Et}+\frac{i}{2\pi}\int_{C'}e^{-iEt}G_{a}(E) dE,
\label{eq:scRs}
\end{eqnarray}
where $C'$ is the new contour shown in Figure~\ref{fig:scRs} and 
$Z^{-1}=[\partial G_a^{-1}(E)/\partial E]_{E=\bar E}$.
Since the imaginary part of $\bar E$ is negative,
the first term, representing the contribution from the simple pole 
on the second Riemannian sheet, yields the exponential decay form. 
As the decay process  proceeds further, the exponential 
term becomes very small and the last integral on the contour $C'$, given by
\andy{intC'}
\beq
\frac{i}{2\pi}\int_{-\infty}^{E_g}e^{-iEt}[G_{a}(E+i\epsilon)-
G_{a}(E-i\epsilon)]dE
\label{eq:intC'}
\eeq
eventually dominates. In the above, 
the second function in the integrand, $G_a(E-i\epsilon)$, must be evaluated 
on the second sheet. Note that $[G_{a}(E+i\epsilon)-G_{a}(E-i\epsilon)]
\propto [\Sigma_{a}(E+i\epsilon)-\Sigma_{a}(E-i\epsilon)]
\propto \lambda^{2}$, where $\lambda$ represents the strength of 
interaction.
The power law can be derived from this expression, under 
appropriate assumptions, as an asymptotic formula for very large $t$. 
This is in accordance with the general theorem (\ref{eq:PaleyWiener}) stated 
before.
The behavior for large $t$, however, depends on the 
details of $\rho_{a}(E')$ and $K_{a}(E')$: For this reason,
in the next subsection, 
we shall examine the above asymptotic formula in the particular case 
discussed in the preceding subsection.

It is important here to mention van Hove's limiting procedure 
concerning the time-scale transformation. 
In spite of their crucial role in the derivation of the closed equations 
(\ref{eq:aUa}) and (\ref{eq:aUadot}) for the survival probability amplitude 
in the particular case studied in Sec.~\ref{sec-longt}.\ref{sec-naive}, 
the conditions (\ref{eq:matrixelement}) on the interaction 
Hamiltonian do not seem quite natural for 
systems with many (ideally infinite) degrees of freedom 
(roughly called ``macroscopic systems"):
The initial state $\vert a\rangle$ (which is now considered to be a direct 
product of a quantum subsystem and the remaining part of the whole system), 
would be given a rather special status if (\ref{eq:matrixelement}) were 
to be satisfied.
On the other hand, the random phase approximation (\ref{eq:nH'n'}), introduced 
at the beginning of this subsection, seems more plausible in such a case.
Pauli\cite{Pauli} first derived a master equation 
in the quantum theory of many-body systems along this line of thought.

Almost three decades later,
van Hove showed that the repeated use of the random phase approximation, 
which is so crucial for the derivation of Pauli's master equation, 
can be avoided by taking the so-called van Hove limit\cite{vanHove} 
\andy{vHlimit}
\beq
\lambda\to0 \quad\mbox{with}\quad\lambda^2t\quad\mbox{kept constant},
\label{eq:vHlimit}
\eeq
where $\lambda$ represents the strength of the interaction, as mentioned above.
He showed that the survival amplitude of some suitably chosen initial 
state decays 
exponentially in this limit, provided that the diagonal-singularity 
assumption holds.
This assumption, which appears reasonable for systems with many degrees 
of freedom, requires the predominance of the diagonal matrix elements 
$\langle\alpha\vert H'H'\vert\alpha\rangle$, which turn out to be of order 
$N$ ($N$ representing the number of degrees of freedom of the system), 
over the remaining off-diagonal terms.
Observe that in our terminology, the limit (\ref{eq:vHlimit}) is equivalent 
to making the time constant $\tau_{\rm G}$, defined in 
Eq.\ (\ref{eq:Gauss}) of Sec.~\ref{sec-stbeh}, and the integral on $C'$ in 
(\ref{eq:scRs}) vanish, for the former is $\sim O(\lambda)$ and the 
latter $\sim O(\lambda^2)t^{-\delta} \,(\delta>0)$
[remember Eq.\ (\ref{eq:intC'}) and the subsequent discussion]:
In conclusion, only the 
exponential decay, characterized just by the lowest-order contribution
$\Sigma_{a}^{(2)}$, remains as a purely probabilistic law.
By means of this procedure, 
van Hove succeeded in deriving the master equation 
from the quantum-mechanical Schr\"{o}dinger equation.
It is also important to notice that in order to realize the exponential law, 
and consequently the master equation, a wave packet must be taken as an 
initial state.
It was shown\cite{iwanami} that consideration of wave packets 
in quantum mechanical scattering processes is equivalent to the extension 
of scattering amplitudes to the complex $E$ plane.
The weak-coupling limit (\ref{eq:vHlimit})
has also been investigated from a mathematical point 
of view.\cite{Davies}$^-$\cite{Accardi}

     Finally, before closing our general discussion, it is worth showing
another method to derive temporal behaviors, without resorting to Fourier or 
Laplace transforms.
It is not difficult to derive the following differential equation for the 
survival amplitude 
\andy{aUIadot}
\barr
& & \!\!\!\!\!\!\!\!\!\!\!\!\!\!\!\!\!\!\!\!\!\!\!\!
 {d\over dt}\langle a\vert U_I(t)\vert a\rangle\nonumber\\
&=&-\int_0^tdt_1\langle a\vert H'_I(t)H'_I(t_1)\vert a\rangle
    \langle a\vert U_I(t_1)\vert a\rangle\nonumber\\
& &+\int_0^tdt_1\int_0^{t_1}dt_2\int_0^{t_2}dt_3
    \langle a\vert H'_I(t)H'_I(t_1)H'_I(t_2)H'_I(t_3)\vert a\rangle_p
    \langle a\vert U_I(t_3)\vert a\rangle\nonumber\\
& &-\int_0^tdt_1\int_0^{t_1}dt_2\int_0^{t_2}dt_3\int_0^{t_3}dt_4
    \int_0^{t_4}dt_5\nonumber\\
& &\quad\times
    \langle a\vert H'_I(t)H'_I(t_1)H'_I(t_2)H'_I(t_3)H'_I(t_4)H'_I(t_5)
    \vert a\rangle_p\langle a\vert U_I(t_5)\vert a\rangle\nonumber\\
& &+\cdots,
\label{eq:aUIadot}
\earr
on the basis of the iterative solution of the Schr\"odinger equation
\andy{recursion}
\barr 
  U_I(t)&=&1-i\int_0^tdt'H_I(t')U_I(t') \nonumber \\
        &=&1-i\int_0^tdt'H'_I(t')
            +(-i)^2\int_0^tdt_1\int_0^{t_1}dt_2H'_I(t_1)H'_I(t_2)
            +\cdots,
\label{eq:recursion} 
\earr 
under the conditions (\ref{eq:nH'n}) (``mass renormalization") and 
(\ref{eq:nH'n'}) (random phase approximation).
Observe that due to these conditions, all terms of odd order give no 
contribution to the survival amplitude.
The ``proper" matrix elements $\langle a\vert\cdots\vert a\rangle_p$ 
in (\ref{eq:aUIadot})
are defined by ($E_{ab}\equiv E_a-E_b$, etc.)
\andy{aac4,aac6}
\barr
&& \!\!\!\!\!\!\!\!\!\!\!\!\!\!\!\!\!\!\!\!\!\!\!\!
\langle a\vert H'_I(t)H'_I(t_1)H'_I(t_2)H'_I(t_3)\vert a\rangle_p\nonumber\\
&\equiv&\sum_{n\not=a}\langle a\vert H'_I(t)H'_I(t_1)\vert n\rangle
                      \langle n\vert H'_I(t_2)H'_I(t_3)\vert a\rangle
                                                               \nonumber\\
&=&\sum_{n\not=a}\sum_{n'\not=a,n}
   e^{iE_{an'}(t-t_3)}e^{iE_{n'n}(t_1-t_2)}
        \vert\langle a\vert H'\vert n'\rangle\vert^2\cdot
        \vert\langle n'\vert H'\vert n\rangle\vert^2,\label{eq:aac4}
\\
\noalign{\vspace*{8pt}}
&& \!\!\!\!\!\!\!\!\!\!\!\!\!\!\!\!\!\!\!\!\!\!\!\!
\langle a\vert H'_I(t)H'_I(t_1)H'_I(t_2)H'_I(t_3)H'_I(t_4)H'_I(t_5)
   \vert a\rangle_p\nonumber\\
&\equiv&\sum_{n\not=a}\sum_{n'\not=a}
        \langle a\vert H'_I(t)H'_I(t_1)\vert n\rangle
        \langle n\vert H'_I(t_2)H'_I(t_3)\vert n'\rangle
        \langle n'\vert H'_I(t_4)H'_I(t_5)\vert a\rangle\nonumber\\
&=&\sum_{n\not=a}\sum_{n'\not=a,n}\sum_{n''\not=a,n,n'}
   e^{iE_{an'}(t-t_5)}e^{iE_{n'n''}(t_1-t_4)}e^{iE_{n''n}(t_2-t_3)}\nonumber\\
& &\phantom{\sum_{n\not=a}\sum_{n'\not=a,n}\sum_{n''\not=a,n,n'}}
   \times
   \vert\langle a\vert H'\vert n'\rangle\vert^2\cdot
   \vert\langle n'\vert H'\vert n''\rangle\vert^2\cdot
   \vert\langle n''\vert H'\vert n\rangle\vert^2,
\label{eq:aac6}
\earr
and so on.
It is evident that these terms correspond to the self-energy parts of
the 4th order $\Sigma_a^{(4)}(E)$, 6th order $\Sigma_a^{(6)}(E)$ and so on, 
in Eq.\ (\ref{eq:SigE}), in Fourier space.

     If ``memory effects" are neglected so that the amplitudes 
$\langle a\vert U_I(t_i)\vert a\rangle$ appearing on the RHS of 
(\ref{eq:aUIadot}) can be evaluated at time $t$ instead of $t_i$, 
the differential equation (\ref{eq:aUIadot})
is solved to give an exponential form for the survival 
amplitude\cite{Tomita}
\andy{aUI(t)a}
\beq
\langle a\vert U_I(t)\vert a\rangle
\simeq e^{-\int_0^t\!d\tau_1(t-\tau_1)f_2(\tau_1)
          +\int_0^t\!d\tau_1\!\int_0^{t-\tau_1}\!d\tau_2
           \!\int_0^{t-\tau_1-\tau_2}d\tau_3
           (t-\tau_1-\tau_2-\tau_3)f_4(\tau_1,\tau_2,\tau_3)+\cdots}.
\label{eq:aUI(t)a}
\eeq
The functions in the exponent are given by
\andy{f2,f4}
\barr
f_2(\tau_1)&=&\langle a\vert H'e^{i(E_a-H_0)\tau_1}H'\vert a\rangle,
\label{eq:f2} \\
f_4(\tau_1,\tau_2,\tau_3)&=&\langle a\vert H'e^{i(E_a-H_0)\tau_1}
                                           H'e^{i(E_a-H_0)\tau_2}
                                           H'e^{i(E_a-H_0)\tau_3}H'
                                                          \vert a\rangle_p,
\label{eq:f4}
\earr
and so on.
These expressions allow us to extract some of the typical temporal 
behaviors of the amplitude (except at very large times).
To this end, let us recall again that the energy spectrum is asymptotically 
very dense in many-body systems and becomes continuous in field-theoretical 
cases so that the summations appearing in (\ref{eq:aac4}) and 
(\ref{eq:aac6}), for example, are to be understood as integrations over 
energy.
Here we consider the following two cases:\cite{Tomita}
\begin{itemize}
\item When the matrix elements of the interaction Hamiltonian $H'$ have a 
      relatively wide energy range over which their variations are small, 
      the phase factors appearing in the functions $f_2,f_4,\ldots$ (see 
      (\ref{eq:f2}) and (\ref{eq:f4})) 
      oscillate rapidly except for very small $\tau$'s.  
      Therefore, after summing over the intermediate states, the 
      functions $f_i$ can be 
      considered to be proportional to $\delta$ functions 
      w.r.t.\ the integration variables $\tau$'s, which results in a 
      linear $t$ behavior
\andy{widebd}
\beq
\langle a\vert U_I(t)\vert a\rangle
\simeq e^{-t\left[
           \int_0^\infty d\tau_1f_2(\tau_1)
           -\int_0^\infty d\tau_1\int_0^\infty d\tau_2\int_0^\infty d\tau_3
           f_4(\tau_1,\tau_2,\tau_3)+\cdots\right]},
\label{eq:widebd}
\eeq
      provided the integrals in the exponent converge.
      Observe that the upper limits of integration can be safely taken 
      equal to $\infty$ under such conditions and therefore the above 
      form becomes a relatively good approximation at large $t$.
      The time $t$ has been assumed to be much larger than the average 
      energy spacing.
\item On the contrary, if the matrix elements have relatively narrow bands, 
      the phase factors in the functions $f_2, f_4,\ldots$ are safely 
      replaced with unity.
      This means that we can replace the arguments of $f_i$ in the 
      integrals by zeros and arrive at
\andy{narrowbd}
\beq
\langle a\vert U_I(t)\vert a\rangle
\simeq e^{-{1\over2}f_2(0)t^2+{1\over4!}f_4(0,0,0)t^4+O(t^6)}.
\label{eq:narrowbd}
\eeq
      Since we have assumed that the time $t$ is small enough so that the 
      above-mentioned replacements are allowed, this approximate form is valid 
      at short times.
      We infer that the dominant term at short times is of second order 
      in $t$, in accordance with the argument presented in the preceding 
      section.
\end{itemize}

     It is worth stressing that the behavior at both short and large times 
has been derived from one and the same expression for the amplitude 
$\langle a\vert U_I(t)\vert a\rangle$ under different physical conditions.
At the same time, however, we should keep in mind that the behavior that 
really dominates in the particular time region under consideration does 
depend on the relative magnitudes of
the Gaussian and the exponential (linear $t$) decays, as well as
on the characteristic time constants of the exponents.

\subsection{Reanalyzing the particular case}
\label{sec-nonexp}
\noindent
In this subsection, we shall reconsider the particular 
interaction Hamiltonian of Sec.~\ref{sec-longt}.\ref{sec-naive} and 
study in detail the analytic properties of the survival amplitude.
This subsection is mainly written in an educational spirit, and  
therefore some parts of it are a specific version of the 
general argument presented in Sec.~\ref{sec-longt}.\ref{sec-generala}.
Nevertheless, we believe that it is very instructive to explicitly
exhibit the temporal behavior at long times, and to perform an analysis
in terms of Laplace transforms.

Let us go back to the definition (\ref{eq:g}) of $g(s,t)$ and see how and 
to which extent the exponential law can be realized.
Since the amplitude $\langle a\vert U_I(t)\vert a\rangle$ is given by the 
inverse Laplace transform (\ref{eq:Laplace}), we understand that the 
exponential decay occurs only when the complex function $g(s,t)$ has 
zeros in the left-half complex $s$ plane.
In order to perform the inverse Laplace transform, the function $g$ has
to be analytically continued into the left-half complex $s$ plane.
By performing the integration over $u$ in (\ref{eq:g}) we obtain
\andy{int.over.u}
\barr
g(s,t)
  &=&s+t\langle a\vert H'{i\over E_a-H_0+i{s\over t}}H'\vert a\rangle
                                                                 \nonumber\\
  &=&s+it\int_{C_0}{\sum_r\vert\langle E_0,r\vert H'\vert a\rangle\vert^2
                             \over E_a-E_0+i{s\over t}}dE_0.
\label{eq:int.over.u}
\earr
Here $\vert E_0,r\rangle$ are eigenstates of $H_0$, and form a complete 
orthonormal set, with $r$ being quantum numbers describing possible 
degeneracies.
Observe that the integrand has a simple pole at $E_0=E_a+i{s\over t}$.
Since $t>0$ and $s$ is taken to have a positive real part (in order to 
assure the convergence of the integration over $t$ in the Laplace 
transform), the pole lies above the 
integration contour $C_0$ which extends along the real $E_0$ axis.
Therefore in order to extend the function $g(s,t)$ into the left-half 
complex $s$ plane, where the real part of $s$ is negative, 
the integration contour 
should be deformed in the complex $E_0$ plane so that its relative 
configuration w.r.t\ the singularity is maintained.\cite{ArakiMKG}\ 
\begin{figure}[htbp]
\vspace*{13pt}
\centerline{\vbox{\hrule width 5cm height0.001pt}}
\vspace*{1.4truein}             
\centerline{\vbox{\hrule width 5cm height0.001pt}}
\vspace*{13pt}
\fcaption{Integration contours (a) $C_0$ for $\Re(s)>0$ and (b) $C_1$ for 
$\Re(s)<0$.
The former extends along the real $E_0$ axis starting from the lowest 
energy $E_g$.
The pole at $E_0=E_a+i{s\over t}$ is located above the contours $C_0$ and
$C_1$.
(c) The contour $C_1$ can further be deformed and decomposed into the 
contour $C_0$ along the real $E_0$ axis and a circle surrounding the 
pole.}
\label{fig:C0C1}
\end{figure}
See Figure~\ref{fig:C0C1}.
This gives us the following change for the last term in (\ref{eq:int.over.u})
\andy{negative.s}
\barr
&&it\int_{C_0}{\sum_r\vert\langle E_0,r\vert H'\vert a\rangle\vert^2
                                 \over E_a-E_0+i{s\over t}}dE_0 \nonumber\\
&&\longrightarrow
 it\int_{C_0}{\sum_r\vert\langle E_0,r\vert H'\vert a\rangle\vert^2
                             \over E_a-E_0+i{s\over t}}dE_0
           +2\pi t\sum_r\vert\langle E_0,r\vert H'\vert a\rangle\vert^2
            \Bigr\vert_{E_0=E_a+i{s\over t}},
\label{eq:negative.s}
\earr
as the pole moves into the lower-half $E_0$ plane crossing the real $E_0$ axis.
Notice that the second term of the RHS, which represents the contribution 
from the simple pole now appearing in the lower-half $E_0$ plane, has 
in general a nonvanishing real part.
We see that this term just expresses the discontinuity of  
the LHS over the imaginary $s$ axis, thus ensuring the analyticity 
of the RHS as a whole.

     It is now clear that the analytically continued function 
$1/g(s,t)$ has the following properties as a function of the complex 
variable $s$:
\begin{itemize}
\item Under the assumption that $H_0$ (like $H$) has a bounded spectrum, 
      a branch 
      cut exists along the imaginary $s$ axis, extending to $-i\infty$ from 
      the branch point at $s=it(E_a-E_g)$, $E_g$ being the lowest 
      value of the spectrum of $H_0$.
\item On the first Riemannian sheet, into which the analytic continuation must 
      be done without crossing the branch cut, the last term 
      in (\ref{eq:negative.s}) does not show up.
      The function $1/g(s,t)$ has no singularity and is analytic in this plane.
\item The last term in (\ref{eq:negative.s}) appears only when the function 
      $g(s,t)$ is continued into the second Riemannian sheet through the cut. 
      This term allows the function $1/g(s,t)$ to have a simple 
      pole in this plane where $\Re(s)<0$.
\end{itemize}

     These properties of the function $1/g(s,t)$ are enough to understand that 
the survival probability amplitude $\langle a\vert U_I(t)\vert a\rangle$, 
being nothing but the inverse Laplace transform of $1/g(s,t)$, exhibits two 
distinct 
behaviors at large times.
Observe that the integration contour over $s$ in (\ref{eq:Laplace}) can be 
deformed in the left-half plane as in Figure~\ref{fig:dcs}(b).
\begin{figure}[htbp]
\vspace*{13pt}
\centerline{\vbox{\hrule width 5cm height0.001pt}}
\vspace*{1.4truein}             
\centerline{\vbox{\hrule width 5cm height0.001pt}}
\vspace*{13pt}
\fcaption{Integration contours over $s$.  
(a) The original one in (\ref{eq:Laplace}) and (b) the deformed one.
The latter is composed of a circle surrounding the simple pole at $s_0$ on 
the second Riemannian sheet and a path starting from $i\infty$ on the second 
sheet, turning around the branch point at $s=it(E_a-E_g)$ and extending back 
to $i\infty$ on the first sheet.}
\label{fig:dcs}\andy{fig:dcs}
\end{figure}
There is a close similarity between the deformed coutours in the Fourier 
(Figure~\ref{fig:scRs}) and in the Laplace transforms 
(Figure~\ref{fig:dcs}).
Let $s_0=-{\gamma\over2}t-i\delta\!Et$ ($\gamma>0$) be the zero of $g(s,t)$ 
in the second Riemannian sheet, so that
\andy{real-part}
\beq
-{\gamma\over2}\Biggl[1+\!
  \int_{E_g}^\infty\!\!{\sum_r\vert\langle E_0,r\vert H'\vert a\rangle\vert^2
  \over\vert E_a-E_0+\delta E-i{\gamma\over2}\vert^2}dE_0\Biggr]
+2\pi\sum_r\vert\langle E_0,r\vert H'\vert a\rangle\vert^2
            \Bigr\vert_{E_0=E_a+\delta E-i{\gamma\over2}}=0,
\label{eq:real-part}
\eeq
\andy{imaginary-part}
\beq
-\delta\!E\Biggl[1-\!
  \int_{E_g}^\infty\!\!{\sum_r\vert\langle E_0,r\vert H'\vert a\rangle\vert^2
  \over\vert E_a-E_0+\delta E-i{\gamma\over2}\vert^2}dE_0\Biggr]
+\int_{E_g}^\infty \!\!{(E_a-E_0)
  \sum_r\vert\langle E_0,r\vert H'\vert a\rangle\vert^2
  \over\vert E_a-E_0+\delta E-i{\gamma\over2}\vert^2}dE_0=0.
\label{eq:imaginary-part}
\eeq
Notice that both $\gamma$ and $\delta E$, which are obtained as solutions of 
these equations, are $t$-independent constants.
Close scrutiny\cite{ArakiMKG} shows that in the weak-coupling limit, 
the second term in the square 
brackets of (\ref{eq:real-part}) gives a finite contribution and we exactly 
reobtain Fermi's Golden Rule (\ref{eq:goldenrule}) with the right factor.
We see again that the presence of the second term in the LHS of 
(\ref{eq:real-part}) is crucial for $g(s,t)$ to have a zero in the left-half 
$s$ plane.
This simple-pole contribution to $1/g(s,t)$ provides us with the exponential
decay form for the survival amplitude, while the integration along the 
imaginary $s$ axis cancels the former to yield the Gaussian 
behavior at short times 
according to the argument exposed in the first part of Sec.~\ref{sec-stbeh}.
As we shall see,
the latter contribution is also responsible for the deviation from the 
exponential law at very long times, as expected on the basis of the
above-mentioned theorem (\ref{eq:PaleyWiener}).

  To estimate the remaining integration along the imaginary $s$ 
axis for very large times, we chose
the integration contour that starts from 
$i\infty$ on the second Riemannian sheet, goes down along the imaginary $s$ 
axis, turns around the branch point at $s=it(E_a-E_g)$, appears into the 
first Riemannian sheet and runs back to $i\infty$ again.
See Figure~\ref{fig:dcs}(b).
The contribution along this contour, which we call $X_C$, is explicitly 
written as\fnm{f}\fnt{f}
{The contribution at the branch point is easily shown to vanish.}
\andy{1st.expression}
\beq
X_C={1\over2\pi i}\int_{itE_{ag}}^{i\infty}\left\{
    {e^s\over g(s,t)}-\left[(s-itE_{ag})\to (s-itE_{ag})e^{-2\pi i}\right]
    \right\}ds.
\label{eq:1st.expression}
\eeq
Here $E_{ag}\equiv E_a-E_g$, and the second term is obtained from the first 
just by replacing $s-itE_{ag}$ with $(s-itE_{ag})e^{-2\pi i}$, 
thus representing the integration on the second Riemannian sheet.
With the change of the integration variable $s=i(y+tE_{ag})$, $X_C$ is 
explicitly written as
\andy{2nd.expression}
\barr
X_C&=&{e^{iE_{ag}t}\over2\pi i}\int_0^{\infty}\Biggl\{
      e^{iy}
      \left[
      y+tE_{ag}+t\int_{E_g}^\infty
             {\sum_r\vert\langle E_0,r\vert H'\vert a\rangle\vert^2
              \over E_g-E_0-{y\over t}}dE_0\right]^{-1}\nonumber\\
  & &\phantom{{e^{iE_{ag}t}\over2\pi i}\int_0^{\infty}\Biggl\{}
      -(y\to ye^{-2\pi i})\Biggr\}dy.
\label{eq:2nd.expression}
\earr

     Let us first calculate the difference between the two integrands.
It arises from the multi-valuedness of the denominator in the integrand, the 
relevant part of which is rewritten, after integration by parts,
\andy{int.by.parts}
\barr
\int_{E_g}^\infty{\sum_r\vert\langle E_0,r\vert H'\vert a\rangle\vert^2
                  \over E_g-E_0-{y\over t}}dE_0
&=&-\ln(E_0-E_g+\mbox{$y\over t$})
    \sum_r\vert\langle E_0,r\vert H'\vert a\rangle\vert^2
    \Bigr\vert_{E_g}^\infty\nonumber\\
& &+\int_{E_g}^\infty\! \ln(E_0-E_g+\mbox{$y\over t$}){d\over dE_0}
    \sum_r\vert\langle E_0,r\vert H'\vert a\rangle\vert^2dE_0\nonumber\\
&\equiv&A.
\label{eq:int.by.parts}
\earr
Notice that the argument of the logarithm is positive because 
$y>0$.
Since the logarithmic function gains an extra 
imaginary part $-2\pi i$ when its 
argument turns  clockwise once around the origin, the above logarithmic 
function changes into
\andy{phase.change}
\beq
\ln(E_0-E_g+\mbox{$y\over t$})\longrightarrow
\ln(E_0-E_g+\mbox{$y\over t$})
-2\pi i\theta(\mbox{$y\over t$}-\vert E_0-E_g\vert)
\label{eq:phase.change}
\eeq
when $y$ is replaced with $ye^{-2\pi i}$.
Therefore by inserting the above expression for the logarithm into formula 
(\ref{eq:int.by.parts}), the denominator in the second term of the RHS in 
(\ref{eq:2nd.expression}) is shown to differ from its counterpart in the 
first term by
\andy{extra.term}
\beq
-2\pi it\sum_r\vert\langle E_g+\mbox{$y\over t$},r
 \vert H'\vert a\rangle\vert^2\equiv -2\pi itB.
\label{eq:extra.term}
\eeq
In this derivation, we have assumed that the 
function $\sum_r\vert\langle E_0,r
 \vert H'\vert a\rangle\vert^2$ vanishes rapidly enough at both boundaries 
in order to make the integration over $E_0$ converge:
\andy{vanbound}
\beq
\sum_r\vert\langle E_0,r\vert H'\vert a\rangle\vert^2\sim
\cases{(E_0-E_g)^\delta&for $E_0\simeq E_g,\quad\delta>0$,\cr
\noalign{\smallskip}
       (E_0)^{-\delta'} &for $E_0\to\infty,\quad\delta'>0$.\cr}
\label{eq:vanbound}
\eeq

     We are now in a position to write down the explicit form of $X_C$.
After a few manipulations, we arrive at ($u=y/E_{ag}t$)
\andy{3rd.expression}
\beq
X_C=-E_{ag}e^{iE_{ag}t}\int_0^\infty
    {B_ue^{itE_{ag}u}du\over[E_{ag}(1+u)+A_u][E_{ag}(1+u)+A_u-2\pi iB_u]},
\label{eq:3rd.expression}
\eeq
where 
\andy{AandB}
\barr
A_u&=&\int_{E_g}^\infty\ln(E_0-E_g+E_{ag}u){d\over dE_0}
      \sum_r\vert\langle E_0,r\vert H'\vert a\rangle\vert^2dE_0,\nonumber\\
B_u&=&\sum_r\vert\langle E_g+E_{ag}u,r\vert H'\vert a\rangle\vert^2.
\label{eq:AandB}
\earr
In the large $t$ limit, i.e., $tE_{ag}\gg1$, the integration over $u$ can be
estimated by the behavior of the integrand for small $u<u_0\sim1/tE_{ag}$.
Recall that the function $B_u$ has been assumed to vanish at the boundary 
$E_0=E_g$, i.e.\ at $u=0$.
As a consequence of (\ref{eq:vanbound}), the function $B_u$ behaves, for 
small u, like
\andy{small.u}
\beq
B_u=\sum_r\vert\langle E_g+E_{ag}u,r\vert H'\vert a\rangle\vert^2
 \sim E_{ag}C_0u^\delta,\qquad\delta>0,
\label{eq:small.u}
\eeq
where the first factor adjusts the dimension of $B_u$, 
while $C_0$ is a positive constant.
We realize that the function $X_C$ exhibits a power decay at 
very large times
\andy{4th.expression}
\barr
X_C&\sim&
   - \; {C_0e^{iE_{ag}t}\int_0^1u^\delta e^{iu}du\over[tE_{ag}]^{1+\delta}
      \left[1+{1\over E_{ag}}\int_{E_g}^\infty\ln(E_0-E_g){d\over dE_0}
      \sum_r\vert\langle E_0,r\vert H'\vert a\rangle\vert^2dE_0\right]^2}
      \nonumber\\
   &\equiv&
   - \; Ce^{iE_{ag}t}{1\over t^{1+\delta}}.
\label{eq:4th.expression}
\earr
Therefore we have shown that the survival amplitude decays as a function of 
the inverse power $t^{-(1+\delta)}$ at very long times, which 
 eventually takes over the exponential behavior.

     To summarize, the survival probability amplitude $\langle a\vert U_I(t)
\vert a\rangle$ is shown to have the following large-time behavior
\andy{large.time}
\beq
\langle a\vert U_I(t)\vert a\rangle\sim
{\cal Z}e^{-{\gamma\over2}t-i\delta Et}-Ce^{iE_{ag}t}{1\over t^{1+\delta}},
\label{eq:large.time}
\eeq
where ${\cal Z}^{-1}=(\partial g(s,t)/\partial s)\vert_{s=-{\gamma\over2}-i
\delta E}$. It is very instructive to 
compare this expression with the general formula (\ref{eq:scRs}) 
of the preceding subsection.
It is evident from the above derivation that, as was already mentioned, 
the very large time behavior 
is in fact governed by the state number density $\rho_a(E)$ and the 
matrix elements $K_a(E)$. See Eq.\ (\ref{eq:Sigint}).
(The correspondence $B\leftrightarrow\rho_aK_a$ is manifest.)

     The power behavior of quantum survival amplitudes at long times 
has attracted the attention of several investigators in the past.
H\"ohler\cite{Hohler} showed explicitly a power behavior 
$\sim t^{-{3\over2}}$ for the Lee model.\cite{Lee}\ 
Knight and Milonni\cite{KnightMilonni'76} investigated 
the two-level Weisskopf-Wigner model of atomic spontaneous emission under 
the dipole and the rotating-wave approximations.
In this model, only two `essential states' are considered,
that is, an excited state and 
no photons present, and the ground state and one photon present.
They calculated the nondecay amplitude using the Laplace transform  
and found a non-exponential tail $\sim t^{-2}$ at long times.
The former behavior corresponds to the 
case $\delta=1/2$ and the latter to $\delta=1$ in 
(\ref{eq:large.time}). It is worth mentioning that these behaviors
 are in complete agreement with our conclusions, 
since the quantity corresponding to the function 
$B_u$ in (\ref{eq:AandB}) is explicitly given and behaves like $\sim\sqrt u$ 
(i.e., $\delta=1/2$) in the former case and $\sim u$ ($\delta=1$) in 
the latter.

\setcounter{equation}{0}
\section{Exponential law in a solvable model}
\label{sec-sdm}
\subsection{The AgBr Hamiltonian}
\label{sec-AgBrH}
\noindent
In order to give a concrete idea of how it is possible 
to obtain the exponential decay law
for a quantum system in interaction with another quantum 
system endowed with a large number of degrees of 
freedom (a ``macroscopic system"), we shall consider the so-called 
AgBr model,\cite{AgBr} that is exactly solvable and has played 
an important role in the quantum measurement problem.
We present this model in the spirit of Havli\v cek's 
remark:\cite{Hav} ``Please try to illustrate your assertion on an example
which would involve $2\times2$ matrices only." Our model involves only
Pauli matrices and some reasonably simple algebra.

We shall focus our attention on the modified version of the AgBr 
model,\cite{NaPa3} that is able to take into account energy-exchange 
processes, and in particular on its weak-coupling, macroscopic 
limit.\cite{NNP} Our exact calculation will 
show that the model realizes the so-called diagonal 
singularity\cite{vanHove} and can display 
the occurrence of an exponential regime {\em at all times}.

The modified AgBr Hamiltonian\cite{NaPa3} describes the interaction 
between an 
ultrarelativistic particle $Q$ and a 1-dimensional $N$-spin array
($D$-system).
The array is a caricature of a linear ``photographic emulsion" of AgBr 
molecules, when one identifies the {\em down} state of the spin with 
the undivided molecule and the {\em up} state with the dissociated 
molecule (Ag and Br atoms). 
The particle and each molecule interact via a spin-flipping local potential.
The total Hamiltonian for the $Q+D$ system reads
\andy{totham} 
\beq
H = H_{0} + H',  \qquad 
\qquad H_0 = H_{Q} + H_{D},
\label{eq:totham} 
\eeq 
where $H_{Q}$ and $H_{D}$, 
the free Hamiltonians of the $Q$ particle and of the ``detector" $D$, 
respectively, and the interaction Hamiltonian $H'$ are written as\fnm{g}\fnt{g}
{In this section, we explicitly write the Planck constant $\hbar$.} 
\andy{H}
\barr 
H_{Q} & = & c \h{p},    \qquad 
     H_{D}  =  \frac{1}{2}  \hbar  \omega
  \sum_{n=1}^{N}  \left( 1+\sigma_{3}^{(n)} \right) , \nonumber  \\
H' & = & \sum_{n=1}^{N} V(\h{x}- x_n)  
  \left[ \sigma_{+}^{(n)} \exp \left( -i \frac{\omega}{c} 
  \h{x} \right) + \sigma_{-}^{(n)} \exp \left( + i \frac{\omega}{c} 
  \h{x} \right) \right], 
\label{eq:H} 
\earr 
where $\h{p}$ is the momentum of the $Q$ particle, $\h{x}$ its position, 
$V$ a real potential, $x_n\; (n=1,...,N)$ the positions of the 
scatterers in the array $(x_n>x_{n-1})$ and $\sigma_{i,\pm}^{(n)}$ 
the Pauli matrices acting on the $n$th site. 
An interesting feature of the above Hamiltonian, as compared to the
original one,\cite{AgBr} is that we are not neglecting 
the energy $H_D$ of the array, namely the energy gap $\hbar \omega$
between the two states of each molecule.
This enables us to take into account energy-exchange 
processes between $Q$ and the spin system $D$.
The original Hamiltonian is reobtained in the $\omega=0$ limit.

The evolution operator in the interaction picture 
\andy{evol} 
\beq 
U_I(t) = e^{iH_{0}t/\hbar} e^{-iHt/\hbar} = 
          e^{-i \int_{0}^{t} H_{I}'(t') dt'/\hbar },
     \label{eq:evol} 
\eeq 
where $H_{I}'(t)$ is the interaction Hamiltonian in the 
interaction picture, can be computed exactly as
\andy{solut} 
\beq
U_I(t) = \prod_{n=1}^{N} \exp \left( -\frac{i}{\hbar} \int_{0}^{t} 
   V(\h{x} + ct'- x_n) dt'
  \left[ \sigma_{+}^{(n)} \exp \left( -i \frac{\omega}{c} 
  \h{x} \right) + \mbox{h.c.} \right]  \right) ,
\label{eq:solut} 
\eeq
and a straightforward calculation yields the $S$-matrix \andy{Smatr} 
\beq 
S^{[N]} = \lim_{t\rightarrow \infty} U_I(t)  = 
\prod_{n=1}^{N} S_{(n)}\ :\qquad  S_{(n)}=
\exp \left( -i\frac{V_{0} \Omega}{\hbar c} 
\mbox{\boldmath $\sigma^{(n)} \cdot u$} \right) ,
\label{eq:Smatr} 
\eeq
where $\mbox{\boldmath $u$} =(
\cos ( \omega x/c), \sin ( \omega x/c), 0 ) $ 
and $V_{0} \Omega \equiv \int_{-\infty}^{\infty} V(x)dx$ is 
assumed to be finite. The above expression enables us to compute 
the ``spin-flip" probability, i.e.\
the probability of dissociating one AgBr molecule: 
\andy{sfprob}
\beq
q = \sin^{2} \left( \frac{V_{0} \Omega}{\hbar c} \right) .    
 \label{eq:sfprob} 
\eeq 

If the initial $D$ state is taken to be the ground state 
$\vert  0 \rangle_N$ ($N$ spins down), and 
the initial $Q$ state is a plane wave $\vert p\rangle$, the final state reads
\andy{Svac}
\beq 
S^{[N]} \vert p, 0 \rangle_N = \sum_{j=0}^N {N\choose j}^{1/2} 
 \left( -i\sqrt{q} \right)^j \left( \sqrt{1-q} \right)^{N-j} 
 \vert p_j, j \rangle_N ,
\label{eq:Svac} 
\eeq 
where $\vert p_j, j \rangle_N$ represents the (spin-symmetrized)
state in which $Q$ has energy $p_j = p -j \hbar \omega/c$ 
and $j$ spins are up.

This enables us to compute several interesting quantities,
such as the visibility of the interference pattern obtained 
by splitting an incoming $Q$ 
wave function into two branch waves (one of which 
interacts with $D$), and the energy ``stored" in $D$ after the interaction 
with $Q$, as well as the fluctuation around the average. The final results 
are
\andy{avener} 
\barr 
{\cal V} & = &  \left(1 - q \right)^{N/2}  
\longrightarrow  e^{- \overline{n} /2}, 
    \nonumber \\ 
\langle H_D \rangle_F & = & qN \, \hbar \omega 
\longrightarrow  \overline{n} \, \hbar \omega , \label{eq:avener} \\ 
\langle \delta H_D \rangle_F & = & 
  \sqrt{\langle \left( H_D - \langle H_D \rangle_F \right)^2 \rangle_F}
  = \sqrt{pqN} \, \hbar \omega  \longrightarrow  \sqrt{\overline{n}} \, 
     \hbar \omega ,  \nonumber 
\earr
where $_F$ stands for the final state (\ref{eq:Svac}), $p=1-q$,
and the trivial trace over the $Q$ particle states
is suppressed. The arrows show the weak-coupling, macroscopic limit 
$N \rightarrow \infty, \; qN = \overline{n}$ = finite.\cite{NaPa3}
All results are exact. It is worth stressing that
$qN=\overline{n}$  represents  the  average  number  of 
excited molecules, so that interference, 
and relative energy fluctuations ``gradually" disappear as
$\overline{n}$ increases.
The limit $N \rightarrow \infty, \; qN < \infty$ is physically very 
appealing, in our opinion, because it corresponds to a {\em finite} energy 
loss of the $Q$ particle after interacting with the $D$ system.
Observe also that (\ref{eq:Svac}) is a generalized 
[$SU(2)$] coherent state and becomes a Glauber coherent state in the
$N \rightarrow \infty, \; qN=$ finite limit. 

\subsection{The exponential law}
\label{sec-explaw}
\noindent
Our next (and main) task is to study the behavior of the propagator,
in order to analyze the temporal evolution of the system.
Observe that the only nonvanishing matrix elements of the 
interaction Hamiltonian $H'$ are those between the eigenstates of $H_0$ 
whose spin-quantum numbers differ by one, so that the conditions 
(\ref{eq:matrixelement}) in Sec.~\ref{sec-longt} are not satisfied.
It is also important to note that the $Q$ 
particle state $\vert cp\rangle$, 
characterized by the energy $cp$, is changed by $H'$ into the state 
$\vert cp\pm\hbar\omega\rangle$, if $\omega\not=0$.
We can therefore expect a dissipation effect and the 
appearance of the diagonal singularity, which leads to the master 
equation.
Following van Hove's pioneering work,\cite{vanHove} one could calculate 
the propagator perturbatively in this model.
However, the solvability of the present model enables us 
to perform a nonperturbative 
treatment and yields an exact expression for the propagator.
Define 
\andy{defalpha} 
\beq 
\alpha_n \equiv \alpha_n(\h{x},t) \equiv
      \int_{0}^{t} V(\h{x} + ct'- x_n) dt' /\hbar ,
\label{eq:defalpha} 
\eeq 
which can be viewed as a ^^ ^^ tipping angle" of the $n$th spin
if one identifies $V$ 
with a magnetic field $B$,\cite{Hiyama} and 
\andy{Smatrag} 
\beq
\sigma^{(n)}_\pm (\h{x}) \equiv
\sigma_{\pm}^{(n)} \exp \left( \mp i \frac{\omega}{c} \h{x} \right) ,
 \nonumber
\eeq
which satisfy, together with $\sigma^{(n)}_3$, the $SU(2)$ algebra
\andy{newalgg}
\beq
\left[ \sigma^{(n)}_- (\h{x}), \sigma^{(n)}_+ (\h{x}) \right]=-\sigma^{(n)}_3,
\qquad
\left[ \sigma^{(n)}_\pm (\h{x}),- \sigma^{(n)}_3 \right]=
\pm 2 \sigma^{(n)}_\pm (\h{x}).
         \label{eq:newalgg}
\eeq
We can now return to the Schr\"odinger picture by inverting 
Eq.\ (\ref{eq:evol}). 
We disentangle the exponential\cite{Bogol}
in $U_I$ by making use of (\ref{eq:newalgg}) and obtain
\andy{relinv}
\beq
e^{-iHt/\hbar} = e^{-iH_{0}t/\hbar} \prod_{n=1}^N
   \left(   e^{-i \tan(\alpha_n) \sigma^{(n)}_+(\hat{x})} 
        e^{-\ln\cos(\alpha_n) \sigma^{(n)}_3}
        e^{-i \tan(\alpha_n) \sigma^{(n)}_-(\hat{x})} \right).
     \label{eq:relinv}
\eeq
Notice that the evolution operators (\ref{eq:solut}) and 
(\ref{eq:relinv}), as well as the $S$-matrix (\ref{eq:Smatr})
are expressed in a factorized form:
This is a property of a rather general class of similar Hamiltonians.\cite{Sun}

We shall now concentrate our attention on the situation in which the $Q$ 
particle is initially located at $x' < x_1$, where $x_1$ is the 
position of the first scatterer in the linear array, and is moving towards 
the array with speed $c$.
The spin system is initially set in the ground state $|0\rangle_N$ of the 
free Hamiltonian $H_D$ (all spins down).
This choice of the ground state is meaningful
from a physical point of view, because the $Q$ particle is initially
outside $D$.

The propagator, defined by  
\andy{propG}
\beq
G(x,x',t) \equiv \, \langle x| \otimes\, _N\langle 0| 
  e^{-iHt/\hbar} |0 \rangle_N \otimes |x' \rangle  ,
     \label{eq:propG}
\eeq
is easily calculated from Eq.\ (\ref{eq:relinv}) and we obtain
\andy{Gstart}
\barr
G(x,x',t) & = & \langle x|\otimes \,
     _N \langle 0| e^{-ic\hat{p}t/\hbar}\prod_{n=1}^N
      \left( e^{-\ln \{ \cos \left[ \alpha_n(\hat{x},t)
       \right] \} \sigma^{(n)}_3} \right) 
       |0 \rangle_N \otimes |x' \rangle \nonumber \\
   & = & \langle x|x'+ct \rangle
      \prod_{n=1}^N
      \left( e^{\ln \{ \cos [\alpha_n(x',t)]\} } \right) 
            \nonumber \\
 & = & \delta (x-x'-ct)
      \prod_{n=1}^N
      \cos \alpha_n(x',t) .
     \label{eq:Gstart}
\earr
Observe that, due to the choice of the free Hamiltonian $H_Q$ in 
(\ref{eq:H}), the $Q$ wave packet does not disperse, and moves 
with constant speed $c$. 
We place the spin array at the far right of the origin ($x_1>0$) and consider
the case where potential $V$ has a compact support and the $Q$ particle 
is initially located at the origin $x'=0$, i.e.\ 
well outside the potential region of $D$.
The above equation shows that the evolution of $Q$ occurs only along 
the path $x=ct$.
Therefore we obtain
\andy{Ginter}
\beq
G(x,0,t) = \delta (x-ct)\prod_{n=1}^N
      \cos \t{\alpha}_n(t) ,\qquad
\t{\alpha}_n(t) \equiv \int_{0}^{ct} V(y - x_n) dy/\hbar c.
        \label{eq:Ginter}
\eeq
This result is {\em exact}. Note that 
\andy{sfproby}
\beq
\sin^{2} \t{\alpha}_n (\infty) =
 \sin^{2} \left( \frac{V_{0} \Omega}{\hbar c} \right) = q 
  \label{eq:sfproby} 
\eeq 
is the spin-flip probability (\ref{eq:sfprob}).
We shall now consider the weak-coupling, macroscopic limit\cite{NaPa3}
\andy{wcmacro}
\beq
q \simeq \left( \frac{V_{0} \Omega}{\hbar c} \right)^2 = O(N^{-1}) ,
 \label{eq:wcmacro}
\eeq 
which is equivalent to the requirement that $\overline{n}=qN$ be finite.
We shall see that the limit is equivalent to the van Hove limit, as given
by Eq.\ (\ref{eq:vHlimit}) of Sec.\ \ref{sec-longt}.\ref{sec-explaw}.
Notice that if we set
\andy{xn}
\beq
x_n = x_1 + (n-1) \Delta,\qquad L=x_N - x_1=(N-1)\Delta,
 \label{eq:xn}
\eeq
the scaled variable $z_n\equiv x_n/L$ can be considered 
as a continuous one $z$ in the above
limit, for $\Delta/L\to0$ as $N\to\infty$.
Therefore, a summation over $n$ is to be replaced by a definite integration
\andy{somma}
\beq
q\sum_{n=1}^Nh(x_n) \rightarrow
q\frac{L}{\Delta}\int_{x_1/L}^{x_N/L}h(Lz)dz
\simeq\overline{n}\int_{x_1/L}^{x_N/L}h(Lz)dz.
    \label{eq:somma}
\eeq
This type of integration gives a finite result if the function $h$ is scale 
invariant, because the integration volume is considered to be finite from 
the physical point of view; in fact, the quantities $x_1/L$ and $x_N/L$ should
be of the order of unity even in the $L\to\infty$ limit.
It will be shown below [Eq.\ (\ref{eq:vaimo})]
that in the present case the function $h$ is 
indeed scale invariant.

For the sake of simplicity, we shall restrict our attention to the case of
$\delta$-shaped potentials, by setting $V(y) = (V_0 \Omega) \delta(y)$. 
This hypothesis is in fact too restrictive:
In the following, we shall see that the requirement that $V$ has a compact
support (local potentials) would suffice.
We obtain 
\andy{vaimo}
\barr
G & \propto & \exp \left( \sum_{n=1}^N \ln \left\{
      \cos \int_{-x_n}^{ct-x_n} (V_0 \Omega/\hbar c) \delta(y) dy 
       \right\} \right)
      \nonumber   \\  
   & = & \exp \left( \sum_{n=1}^N \ln \left\{
      \cos \left[ (V_0 \Omega/\hbar c) \theta(ct-x_n) \right] \right\} \right)
      \nonumber   \\  
   & \rightarrow & \exp \left( - \frac{\overline{n}}{2} 
      \int_{x_1/L}^{x_n/L} \theta(ct-Lz) dz \right)
      \nonumber   \\  
   & = & \exp \left( - \frac{\overline{n}}{2} \left[
      \frac{ct-x_1}{L}  \theta(x_N-ct)\theta(ct-x_1)
      + \theta(ct-x_N) \right]  \right),
   \label{eq:vaimo}   
\earr
where the arrow denotes the 
weak-coupling, macroscopic limit (\ref{eq:wcmacro}).

This brings about an exponential regime {\em as soon as the interaction
starts}:
Indeed, if $x_1 < ct < x_N$,
\andy{explaw}
\beq
G \propto \exp \left( - \overline{n} 
      \frac{c(t-t_0)}{2L} \right),
   \label{eq:explaw}   
\eeq
where $t_0 = x_1/c$ is the time at which the $Q$ particle meets the first
potential.
On the other hand, if $ct > x_N$ (that corresponds to the case in which $Q$ 
has gone through $D$ and the interaction is over) we have
\andy{tinflim}
\beq
G \propto e^{-\overline{n}/2}.
   \label{eq:tinflim} 
\eeq
This could also be obtained directly from (\ref{eq:Ginter})
and is in complete agreement with the first formula in (\ref{eq:avener}),
because $|G|^2$ is nothing but the probability that $Q$ goes
through the spin array {\em and} leaves it in the ground state.

Notice that there is {\em no} Gaussian behavior at short times.
As was discussed in Sec.~\ref{sec-stbeh} and Ref.\ \citelow{lowbnd},
deviations from the 
exponential behavior at short times are a consequence of the finiteness of 
the mean energy of the initial state. 
Observe that the mean energy of the 
position eigenstates  is not well-defined in this model.  
If the position eigenstates in 
Eq.~(\ref{eq:propG}) are substituted with wave packets of size $a$,
the mean energy aquires a well-defined meaning.
It is shown below that in this case the exponential regime is attained a short 
time after $t_0=x_1/c$, of the order of $a/c$.

Let $\vert\tilde0\rangle$ be a wave packet of size $a$, initially distributed 
around the origin $x'=0$
\andy{wp0}
\beq
\vert\tilde0\rangle=\int dxC(x)\vert x\rangle\otimes\vert0\rangle_N.
\label{eq:wp0}
\eeq
For simplicity, we may choose the following expression for the wave packet 
\andy{C(x)}
\beq
C(x)={1\over\sqrt a}\theta(a/2-\vert x\vert) e^{ip_0x/\hbar}.
\label{eq:C(x)}
\eeq
(One should choose a smooth $C$ function in order to avoid 
singularities, however, in such a case the final expression 
would be slightly more involved and nothing essential would change.)
We assume here that the initial wave packet has no overlap with the potential 
region $[x_1,x_N]$ and that the size of the former is much smaller
than that of the latter
\andy{sizes}
\beq
a/2<x_1,\qquad a\ll L=x_N-x_1.
\label{eq:sizes}
\eeq
Observe that the mean energy of the initial state is finite, for
one obtains
\andy{menergy}
\beq
\langle\tilde0\vert c\h p\vert\tilde0\rangle
= c p_0 < \infty .
\label{eq:menergy}
\eeq
We calculate the state vectors 
\andy{psis}
\beq
\vert\psi(t)\rangle=e^{-iHt/\hbar}\vert\tilde0\rangle,\qquad
\vert\psi_0(t)\rangle=e^{-iH_0t/\hbar}\vert\tilde0\rangle,
\label{eq:psis}
\eeq
evolved under the action of 
the total Hamiltonian $H$ and the free Hamiltonian $H_0$, 
respectively, and define the propagator
\andy{newG}
\beq
G(t)\equiv\langle\psi_0(t)\vert\psi(t)\rangle,
\label{eq:newG}
\eeq
which extracts the net effect of the interaction.
A straightforward calculation shows that the propagator $G$ has the following 
temporal behavior in the weak-coupling, macroscopic limit (\ref{eq:wcmacro})
\andy{explicitG}
\beq
G(t)=\cases{1            &for $0<ct<x_1-a/2$,\cr
\noalign{\smallskip}
            g_1(t)       &for $x_1-a/2<ct<x_1+a/2$,\cr
\noalign{\smallskip}
            g_2(t)       &for $x_1+a/2<ct<x_N-a/2$,\cr
\noalign{\smallskip}
            g_3(t)       &for $x_N-a/2<ct<x_N+a/2$,\cr
\noalign{\smallskip}
            e^{-\bar n/2}&for $x_N+a/2<ct$,\cr}
\label{eq:explicitG}
\eeq
where the functions $g_i(t)$ ($i=1,2,3$) are given by
\andy{gi}
\barr
g_1(t)&=&1+{2L\over\bar n a}
         \left[1-\exp\left(-{\bar n\over2L}(ct-x_1+a/2)\right)\right]
          -{1\over a}(ct-x_1+a/2),\nonumber\\
g_2(t)&=&{2L\over\bar n a}\left[ 1-e^{-\bar n a/2L} \right]
         \exp\left(-{\bar n\over2L}(ct-x_1-a/2)\right), \label{eq:gi} \\
g_3(t)&=&e^{-\bar n/2}
         \left\{
          1+{2L\over\bar n a}
            \left[\exp\left(-{\bar n\over2L}(ct-x_N-a/2)\right)-1\right]
          +{1\over a}(ct-x_N-a/2)\right\}. \nonumber 
\earr
See Figure 7.
\begin{figure}[htbp]
\vspace*{13pt}
\centerline{\vbox{\hrule width 5cm height0.001pt}}
\vspace*{1.4truein}             
\centerline{\vbox{\hrule width 5cm height0.001pt}}
\vspace*{13pt}
\fcaption{Temporal behavior of the propagator $G(t)$ defined in 
(\ref{eq:newG}). See (\ref{eq:explicitG})}
\end{figure}

We realize that only the function $g_2(t)$ exhibits the exponential
form and that in the time intervals $x_1/c-a/2c<t<x_1/c+a/2c$ and 
$x_N/c-a/2c<t<x_N/c+a/2c$, the propagator $G$ does not 
follow an exponential behavior.
In the first interval the wave packet has not yet completely entered 
the potential region and the interaction between the $Q$ particle and 
the spin system $D$ has not yet attained its full strength.
Analogously, in the second region, the wave packet is going out of the array.
Notice that in this model the exponential law holds if the whole wave packet 
is merged in the potential region so that the interaction keeps its full 
strength: See $g_2(t)$ in (\ref{eq:gi}).
This suggests an equivalent, maybe more ``suggestive" interpretation:
one could say that in the first region $x_1/c-a/2c<t<x_1/c+a/2c$,
the ``initial state" has not been ``prepared", yet. 
(By ``initial", in the above, we do {\em not} mean an eigenstate of the free
Hamiltonian, but rather, intuitively, the state 
that will undergo the exponential decay.)
This alternative viewpoint sheds some light 
on the importance of what is usually referred to as ``state preparation"
in quantum mechanics: It is {\em first} necessary to {\em prepare}
the right initial state, in order to observe a certain temporal 
(e.g.\ exponential) development of this state.

The region $x_1/c-a/2c<t<x_1/c+a/2c$ may be viewed 
as a possible residuum of the short-time Gaussian behavior.
In fact, in this region, the function $g_1(t)$ behaves like 
\andy{g1}
\beq
g_1(t)\sim1-{1\over4\bar naL}(ct-x_1+a/2)^2 ,
\label{eq:g1}
\eeq
and a similar behavior of $g_3(t)$ is found in the final region
of the linear array.
It is worth stressing that the temporal behaviors obtained in 
(\ref{eq:explaw}) and in (\ref{eq:explicitG}) are in complete agreement with 
some general theorems.\cite{nm,FGR,Misra}

The absence of the power behavior at very long times in 
this model may be traced back to the special character of our free 
Hamiltonian $H_Q$ in (\ref{eq:H}):
The spectrum of the operator $c\h p$ is {\em not} bounded from below,
so that the total Hamiltonian $H$ does not possess a lower-bounded spectrum.
Of course we can safely consider, on physical grounds, only those 
incident $Q$ particles that have enough (positive) energy to go through the 
spin system, in order to assure the positivity of the energy of the total 
system. Nevertheless, it should be 
recalled that the boundedness of spectrum of the total Hamiltonian is the 
only condition required for the Paley-Wiener theorem 
(\ref{eq:PaleyWiener}) in Sec.~\ref{sec-longt}.
Our model system is out of the range of applicability of the theorem and 
therefore its exponential behavior at very long times is 
not in contradiction with it.
Similarly, no ground state of the free Hamiltonian $H_0$ in 
(\ref{eq:totham}) exists owing to the presence of the $c\h p$ operator and 
the analysis developed in Sec.~\ref{sec-longt}.\ref{sec-nonexp} is not 
applicable to this case either.
In this respect, it would be interesting to consider the lower-bounded 
Hamiltonian $H_Q = c |\h p|$, but, unfortunately, this operator raises 
serious problems that are closely related to the early atttempts at 
defining a time operator in quantum mechanics.\cite{Paulitime}

One might suspect that the approximation of $\delta$-shaped potential is 
playing an important role in the derivation of the exponential law.
As a matter of fact, this is not the case:
A detailed calculation, making use of square potentials
of strenght $V_0$ and width $\Omega$, centered at each $x_n$, yields, for 
$x_1+\frac{\Omega}{2} < ct < x_N-\frac{\Omega}{2}$,
\andy{quasiexp}
\beq
G \propto \exp \left( - \overline{n} 
      \frac{ct-x_1}{2L} + \frac{\overline{n}\Omega}{12L}  \right).
   \label{eq:quasiexp}   
\eeq
In this case, the exponential regime is attained a short time after $t_0$,
of the order of the width of the potential $V$, which, in the present
model, can be made arbitrarily small.
Also in the present context, the region $t\sim t_0+O(\Omega/c)$ may be 
viewed as a residuum of the short-time Gaussian behavior.
However, it is clear from Eq.\ (\ref{eq:quasiexp}) that once the particle
has {\em completely} penetrated into the potential region, the exponential
law holds {\em exactly} in the weak-coupling, macroscopic limit.
This is, in our opinion, one of the most interesting features of the AgBr 
Hamiltonian.

It is very interesting to bring to light the 
profound link between the weak-coupling, macroscopic
limit $qN = \overline{n}=$ finite considered in this
model and van Hove's ^^ ^^ $\lambda^2 T$" limit.\cite{vanHove}
First, it is important to note that in the weak-coupling, macroscopic limit,
van Hove's diagonal singularity naturally appears in the 
present model.
It is easy to check that for each diagonal matrix element of $H'^2$,
there are $N$ intermediate-state contributions:
Indeed, for example
\andy{hsqu}
\beq
\langle0,\ldots,0\vert H'^2\vert0,\ldots,0\rangle
=\sum_{j=1}^N\vert\langle0,\ldots,0\vert H'\vert0,\ldots,0,1_{(j)},0,
                   \ldots,0\rangle\vert^2.
   \label{eq:hsqu}   
\eeq
On the other hand, at most two states can contribute 
to each off-diagonal matrix element of $H'^2$.
This ensures that only the diagonal matrix elements are kept in the 
weak-coupling, macroscopic limit, $N\to\infty$ with $qN<\infty$, which is 
the realization of diagonal singularity in our model.

The link between the weak-coupling, macroscopic
limit $qN = \overline{n}=$ finite considered above 
and van Hove's ``$\lambda^2 T$" limit
can be also intuitively justified as follows: 
The free part of the particle Hamiltonian is 
$H_{Q}= c\h{p}$, so that the particle travels with constant speed
$c$, and interacts with the detector during the total time
$T = L/c \simeq N\Delta/c$. 
Since the coupling constant $\lambda \propto V_0 \Omega$,
one gets $\lambda^2 T \propto (V_0 \Omega)^2 N \Delta/c \propto qN$.
Notice that the ``lattice spacing" $\Delta$, the inverse of which 
corresponds to a density in our 1-dimensional model, can be kept finite 
in the limit. 
(In such a case, we have to express everything in terms of scaled 
variables, that is, $\tau\equiv t/L$, $z_1$ and $z_N$, introduced just after 
Eq.~(\ref{eq:xn}) and $\zeta\equiv a/L$, where $a$ is the size of the wave 
packet.)

The role played by the 
energy gap $\omega$ need also be clarified.
First of all, $\omega$ plays a very important role to guarantee 
the consistency of the physical framework: 
If $\omega=0$, all spin states would be energetically 
degenerated and the choice of 
the $\sigma_3$-diagonal representation would be quite {\it arbitrary\/}.
In other words, only a nonvanishing $\omega$ (or $H_D$) logically enables us 
to use the eigenstates of $\sigma_3$ in order to evaluate 
the relevant matrix elements. 
Although $\omega$ does not appear in our final results (\ref{eq:explaw}) or 
(\ref{eq:explicitG}), it certainly does in other propagators,
involving an initial state of the spin system different from
$|0 \rangle_N$.

Finally, we would like to shortly comment on other possible 
causes for the occurrence of the exponential behavior
displayed by our model. This is a delicate problem.
Our analysis suggests that the exponential behavior
(\ref{eq:explaw}) is mainly due to the locality of the potentials $V$ and the
factorized form  of the evolution 
operator (\ref{eq:relinv}), which shows that the interactions
between $Q$ and adjacent spins of the array are independent, and the evolution 
``starts anew" at every step. This suggest the 
occurrence of a sort of Markovian process, which would be in
agreement with the purely dissipative behavior (\ref{eq:explaw}).
A preliminary calculation shows that the operator 
\andy{wiecorr}
\beq
\h{W} (t) = c\h{p} (t) - \langle c\h{p} (t) \rangle ,
   \label{eq:wiecorr}   
\eeq
where $\h{p}(t)$ is in the Heisenberg picture and 
the average is taken over a wave packet of $Q$ and the ground state of $D$,
has several interesting features: In particular, some properties of 
$\h{W}(t)$ closely resemble those of a Wiener process.
In this context, the connection between the exponential ``probability
dissipation" (\ref{eq:explaw}) and the (practically irreversible)
energy-exchange between the $Q$ particle and the ``environment" $D$ 
is a very open problem and should be investigated in detail.
Leggett's remark,\cite{Leggett} 
about the central relevance of the problem of 
dissipation to decoherence and
quantum measurement theory make the above topic 
very interesting: There is, in our opinion, a profound link between
decoherence, irreversibility, ``probability dissipation" and genuine energy
dissipation. This link is not yet fully understood.
Work is in progress to clarify the above issues.

\setcounter{equation}{0}
\section{Conclusions and additional comments}
\label{sec-concom}
\noindent
We have analyzed the temporal behavior of quantum mechanical systems at 
all times, by taking into account seminal as well as more recent results.
As we have seen, the temporal development of the survival probability
of a suitable initial state undergoes an evolution that can be roughly 
decomposed into three regions: A Gaussian region at short times,
an approximately exponential region at intermediate times and 
a power region at very long times.

The short-time Gaussian behavior is ascribable to a very general property
of the temporal development engendered by the Schr\"odinger equation.
As we have seen, this leads to the quantum Zeno effect and has 
very interesting spin-offs on the quantum measurement problem. In particular, 
we have argued that the QZE is by no means a
proof in support of the Copenhagen 
interpretation, based on von Neumann's projecion postulate: Indeed, a purely 
dynamical explanation of the QZE 
was put forward, completely based on unitary evolutions.
We have also discussed the important difference between the QZE and the QZP,
and have argued that the latter is in contradiction with Heisenberg's 
uncertainty principle.

One of our primary concerns has been to understand
under which physical conditions we can observe an exponential behavior, 
since, as was already stressed, no deviation from the exponential law has ever 
been observed.\cite{YY93b}
The exponential form is representative of a totally 
incoherent nature of the system under consideration.
Remember that the notion of coherence is one of the essential features of the
quantum world, where the evolution (of the total system) is always unitary.
When we focus our attention on one part (subsystem) of the total system,
if the latter is endowed with a huge number of degrees of freedom
and can be treated as a kind of reservoir, we can expect the former to behave
in a stochastic manner.
The interaction between the subsystem and the remaining part of the total
system may, under such circumstances, cause dissipation on the former.
It is then natural to expect the appearance of irreversible 
phenomena, governed by master equations, for such subsystems.
This is, however, by no means a trivial matter:
Because quantum dynamics, i.e.\ the Schr\"odinger equation,
is symmetric in time, we obviously expect the survival probability 
to go back to unity
after a certain elapse of time, which is of the order of Poincar\'e time.
However, a purely exponential decay of the amplitude would actually 
imply an actual (``eternal") loss of probability, that would never
be recovered.
Rigorously speaking, this is at variance with the underlying 
unitarity of the temporal evolution.
An appropriate limiting procedure is required in order to explicitly derive 
the exponential law, as was repeatedly remarked.

Since it is {\em mathematically} true that quantum systems never behave 
exponentially at very long times, 
it is clear that in order to realize the exponential law at long 
times on the basis of quantum theory, some 
additional manipulation is necessary. By considering 
the exponential law as a manifestation 
of the classical character of the system, we can recognize the importance of 
such a notion as the transition from a quantum to a classical regime.

Fermi's Golden Rule,\cite{Fermigold}
according to which we obtain the decay constant 
(\ref{eq:goldenrule}) in Sec.~\ref{sec-longt},
is a lowest-order approximation.
We believe that the quantum mechanical lowest-order perturbation theory 
cannot be considered completely satisfactory, in this context,
unless it is appropriately supported by some additional limiting 
procedure that accounts for the macroscopic nature of the
total system with which the decaying system is interacting.

We have also seen that van Hove's procedure, based on the 
weak-coupling, macroscopic limit considered in Sec.\ \ref{sec-longt},
yields a satisfactory answer to many of the above problems.
In particular, his procedure dispenses 
with Pauli's random phase assumptions
and yields an exponential behavior at all times, by cancelling
both the short-time Gaussian region and the long-time power tail.
Of course, one is led to wonder about the meaning of the 
``$\lambda^2 T$" limit, and in particular about the very reasons why 
Nature should follow this prescription. Even though no clear-cut answer can 
be given to this very deep question, we believe that there is much more to 
be understood, on this issue.

What we are actually looking for is a consistent way of deriving the classical
nature (loss of coherence) from quantum theory in a certain limit which 
should properly characterize the classical and macroscopic nature.
At this point one may see a close connection with the quantum measurement 
problem,\cite{von}$^-$\cite{Bush,np} 
where the realization of the classical nature of experimental results, namely 
the appearance of exclusive events after a
measurement, has been one of the central 
issues of controversy since the birth of quantum mechanics.

As a general rule, explicit examples can be often
more useful than general abstract considerations.
As Berry recently put it, 
``Only wimps specialize in the general case. Real scientists pursue 
examples".\cite{Berry}
In the attempt to follow the same, general philosophy, 
we have endeavoured to clarify the above issues by considering a
particular solvable model, known as modifed AgBr Hamiltonian:
This model enjoys several interesting features:
It is solvable and relatively simple. It also yields very useful
insights into general topics like quantum measurements and dissipation.
Finally, it discloses the occurrence of van Hove's limit in a very
direct way.

As we have seen, if the initial state of the spin array 
in the AgBr model is chosen to be 
the ground state of the free Hamiltonian, in which all spins are down,
the survival probability of such a state, once the interaction with an
incoming external particle has been switched on, behaves in
a purely exponential way. The causes of this behaviour,
as well as the physical approximations they reflect, have been 
analyzed in detail, also in the light of general theorems.

In conclusion, we believe that further understanding in all the above-mentioned
topics can be achieved by analyzing the general structure 
of quantum mechanical evolutions and, at the same time, by 
looking at interesting models and particular limiting procedures.

\nonumsection{Acknowledgements}
\noindent We acknowledge interesting discussions with Prof.\ I.\ Ohba.
M.N.\ was partially supported by the Japanese Ministry of Education,
Science and Culture,
and S.P.\ by the Japanese Society for the Promotion of Science, under a
bilateral exchange program with Italian Consiglio Nazionale 
delle Ricerche, and by the Administration Council of the University of Bari.
S.P. thanks the High Energy Physics Group of Waseda University
for their very kind hospitality.

\nonumsection{References}

\end{document}